\documentclass[apj]{emulateapj}

\newcommand{\kms}{\,km~s$^{-1}$}

\newcommand{\Msun}{\mbox{\,$M_{\odot}$}}
\newcommand{\Lsun}{\mbox{\,$L_{\odot}$}}
\slugcomment{To be published in ApJ}

\usepackage{lscape}
\usepackage{natbib}
\usepackage{url}
\font\smcap=cmcsc10

\newcommand{\degree}{$^\circ$}
\newcommand{\nai}{Na\,{\smcap i}}
\newcommand{\caii}{Ca\,{\smcap ii}}

\newcommand{\vio}{$(V-I)_0$}
\newcommand{\ivi}{($I,\,V-I$)}
\newcommand{\ivio}{[$I_0,\,(V-I)_0$]}

\newcommand{\feh}{$\rm[Fe/H]$}
\newcommand{\fehp}{$\rm[Fe/H]_{phot}$}
\newcommand{\fehs}{$\rm[Fe/H]_{spec}$}

\newcommand{\vhel}{$v_{\rm hel}$}

\newcommand{\olike}{$\langle L_i\rangle$}

\newcommand{\mvsph}{$\langle v\rangle^{\rm sph}$}
\newcommand{\sigvsph}{$\sigma^{\rm sph}_v$}

\newcommand{\rproj}{$R_{\rm proj}$}
\newcommand{\Gsph}{$G^{\rm sph}(v)$}

\shorttitle{M31's Giant Southern Stream}
\shortauthors{Gilbert et~al.}

\begin{document}
\bibliographystyle{apj}

\title{A Spectroscopic Portrait of Andromeda's Giant Southern Stream}

\author{
Karoline~M.~Gilbert\altaffilmark{1,2},
Puragra~Guhathakurta\altaffilmark{2},
Priya Kollipara\altaffilmark{2},
Rachael~L.~Beaton\altaffilmark{3},
Marla~C.~Geha\altaffilmark{4},
Jason~S.~Kalirai\altaffilmark{5},
Evan~N.~Kirby\altaffilmark{6,2},
Steven~R.~Majewski\altaffilmark{3},
and Richard~J.~Patterson\altaffilmark{3}
}

\email{
kgilbert@astro.washington.edu}

\altaffiltext{1}{Department of Astronomy, University of Washington, Box 351580, Seattle, WA, 98195-1580.}
\altaffiltext{2}{UCO/Lick Observatory, Department of Astronomy \&
Astrophysics, University of California Santa Cruz, 1156 High Street, Santa
Cruz, California 95064.}
\altaffiltext{3}{Department of Astronomy, University of Virginia, P.O.\
Box~400325, Charlottesville, VA 22904-4325.}
\altaffiltext{4}{Astronomy Department, Yale University, P.O.\ Box 208101, New Haven, CT 06520-8101.}
\altaffiltext{5}{Space Telescope Science Institute, 3700 San Martin Drive, Baltimore, MD 21218.}
\altaffiltext{6}{Astronomy Department, California Institute of Technology, 1200 East California Blvd, 
Pasadena, CA 91125.}
\setcounter{footnote}{1}

\begin{abstract}
The giant southern stream (GSS) is the most prominent tidal debris feature in
M31's stellar halo and covers a significant fraction of its southern
quadrant.  The GSS is a complex structure composed of a relatively
metal-rich, high surface-brightness ``core'' and a lower metallicity, lower
surface brightness ``envelope.''  We present spectroscopy of red giant stars
in six fields in the vicinity of M31's GSS (including four new fields and
improved spectroscopic reductions for two previously published fields) and
one field on Stream~C, an arc-like feature seen in star-count maps on M31's
southeast minor axis at $R\sim60$~kpc.  These data are part of our on-going
SPLASH survey of M31 using the DEIMOS instrument on the Keck~II 10-m telescope.
Several GSS-related findings and measurements are presented here.  We
present the innermost kinematical detection of the GSS core to date
($R=17$~kpc).  This field also
contains the inner continuation of a second kinematically cold component that was
originally seen in a GSS core field at $R\sim21$~kpc.  The velocity 
gradients of the GSS and the second component in the combined data set are
parallel over a range of $\Delta{R}=7$~kpc, suggesting that this may
represent a bifurcation in the
line-of-sight velocities of GSS stars.  We present the first kinematical
detection of substructure in the GSS envelope (S quadrant, $R\sim58$~kpc).
Using kinematically identified samples, we show that the envelope debris has
a $\sim0.7$~dex lower mean photometric metallicity and possibly higher
intrinsic velocity dispersion than the GSS core.  The GSS is also 
identified in the field
of the M31 dwarf spheroidal satellite And~I; the GSS in this field has a metallicity distribution
identical to that of the GSS core.  We confirm the previous finding of two
kinematically cold components in Stream~C, and measure intrinsic velocity
dispersions of $\sim10$ and $\sim4$~km~s$^{-1}$.
This compilation of the kinematical (mean velocity, intrinsic
velocity dispersion) and chemical properties of stars in the GSS core and
envelope, coupled with published surface brightness measurements and
wide-area star-count maps, should improve constraints on the orbit and
internal structure of the dwarf satellite progenitor.

\end{abstract}

\keywords{galaxies: halo --- galaxies: individual (M31) --- stars: kinematics --- techniques:
spectroscopic}

\setcounter{footnote}{0}

\section{Introduction}\label{sec:intro}
Galaxy mergers play a key role in galaxy formation and evolution \citep[e.g.,][]{searle1978,white1978,abraham1996,springel2005}.  Observations have uncovered abundant evidence of the ubiquity of minor mergers in the form of tidal streams.  Those discovered in our own Milky Way (MW) include the Magellanic stream \citep{mathewson1974}, Monoceros stream \citep{yanny2003,rocha-pinto2003}, and Sagittarius Stream \citep{ibata1994,majewski2003,newberg2003}.  
Detailed studies of tidal debris features can provide insight into statistical models of hierarchical structure formation, the structure and merger history of an individual galaxy, and the properties of dwarf satellite systems.  Due to their sparse stellar density and long
dynamical times, stellar halos are ideal for investigating the merger history
of an individual galaxy in detail as it is possible for higher density 
tidal debris features to remain identifiable in phase-space for 
billions of years.  

Observations of the Galaxy and
theoretical arguments indicate that present-day dwarf satellite galaxies are
different from the earlier generations of dwarf galaxies that merged to form the bulk of the virialized stellar halo.  Stars 
in the classical satellites of the Galaxy on average have lower relative 
abundances of alpha elements than stars of similar 
metallicities in the Galaxy's halo 
\citep{fuhrmann1998,shetrone2001,shetrone2003,venn2004,geisler2007},
although the relative abundances of alpha elements are similar in the lowest 
metallicity stars \citep{kirby2009,frebel2009}.
This has been shown to be a 
natural consequence of the hierarchical formation framework: simulations 
of stellar halo formation indicate that the majority of stellar mass in 
the halo was contributed by 
massive satellite galaxies that were accreted and disrupted early on, while 
surviving satellites are typically accreted much 
later (resulting in more metal-enrichment via Type Ia 
supernovae) and are on average less massive  (and experienced less efficient 
star formation) than the average halo building block \citep{robertson2005,font2006apjsats}.  

Minor mergers caught in the 
process of disruption bridge the
gap between the surviving and completely disrupted dwarf populations.  The giant southern stream \citep[GSS;][]{ibata2001nature} in the Andromeda Galaxy (M31) offers a rare glimpse at a recent, significant
accretion event in an external galaxy that is close enough \citep[$d_{\rm M31}=783$~kpc;][]{stanek1998,holland1998} to allow spectroscopy of
individual red giant branch (RGB) stars.  This stream is believed to result from the merger with M31 of a
galaxy of stellar mass $\sim 2\times10^9$~\Msun\ less than 1~Gyr ago \citep{fardal2007}, 
and is estimated to have contributed at least 10\% of the stars in M31's halo \citep[assuming azimuthal symmetry of M31's halo;][]{font2006gss,ibata2007}.  

Ground-based photometric observations have shown that the GSS pollutes a large portion of the southern quadrant of M31's stellar halo.  It is observable to a projected distance of \rproj\,$\sim 100$~kpc from the center of M31, with a metal-rich ``core''\footnote{In this context ``core'' refers to the high surface 
brightness portion of the GSS extending along its length, not the GSS's 
progenitor, which has yet to be identified.} surrounded by a lower metallicity ``envelope'' \citep{ibata2007}. 
The line-of-sight distance to the GSS has been measured from \rproj\,$\sim 10$ to 60~kpc using the apparent magnitude of the tip of the RGB. The GSS is located $\sim 100$~kpc behind M31 at \rproj\,$\sim 60$~kpc, but is at the same distance as M31 at  \rproj\,$\sim 10$~kpc \citep{mcconnachie2003}.  The detailed star formation history of a field located on the GSS has been derived from ultra-deep HST observations that reach below the main sequence turn-off; the mean age of the stellar population is 8.8~Gyr (70\% of the stars are younger than 10~Gyr), and the mean metallicity is $-0.7$~dex \citep{brown2006apj}.

Spectroscopic observations of RGB stars in M31's GSS have been limited to the  metal-rich core. Line-of-sight velocity distributions have been published for fields ranging from \rproj\,$=12$ to 60~kpc in projected radial distance from M31's center \citep{ibata2004,guhathakurta2006,kalirai2006gss}.  At \rproj\,$=60$~kpc, the GSS's mean line-of-sight (heliocentric) velocity is \vhel\,$=-320$~\kms.   The innermost unambiguous detection of the GSS is in a field at \rproj\,$=21$~kpc, with a mean line-of-sight velocity that is blueshifted substantially relative to the 60~kpc field: \vhel\,$=-513$~\kms\ \citep{kalirai2006gss}.  The 12~kpc field does not have a secure GSS detection; only 2\,--\,4 stars are found that may be related to the GSS based on their velocities \citep{ibata2004}.  In each field in which it has been observed, the GSS has a narrow velocity dispersion of $\sim 10$ to 20~\kms.   

These existing photometric and spectroscopic data sets have inspired detailed models of the collision that produced M31's GSS.  It is believed to have formed following a close pericentric passage of an M31 satellite on a highly radial orbit \citep{ibata2004,font2006gss,geehan2006,fardal2006,fardal2007,mori2008}.  The \citet{fardal2007} numerical models showed that several
other observed substructures in M31 [the Northeast (NE) and Western (W)
shelves] can be explained as the forward continuation of the GSS, and predict a third SE shelf feature that has since been
discovered spectroscopically \citep{gilbert2007}.  The latest models demonstrate that several of the observed properties of the GSS and other debris in M31, including the asymmetric surface brightness distribution of the GSS \citep{mcconnachie2003}, the metal-rich core and metal-poor envelope structure \citep{ibata2007}, and perhaps even the minor-axis arcs \citep{ibata2007}, can be explained by a rotating progenitor with a reasonable initial metallicity gradient \citep{fardal2008}.  Despite these recent successes, the current location of the progenitor's core and whether or not it is still intact remain unknown.  More observations of the GSS and related debris are needed to further constrain models of the collision and produce more precise predictions for the current location and dynamical state of the progenitor's core.  

As part of the Spectroscopic and Photometric Landscape of Andromeda's Stellar Halo (SPLASH) survey, we have conducted an extensive spectroscopic survey of M31 RGB stars using the DEIMOS spectrograph on the Keck~II telescope \citep{guhathakurta2005,kalirai2006halo,gilbert2006,gilbert2007}, including observations of the GSS \citep{guhathakurta2006,kalirai2006gss}.  In this contribution, we present spectroscopic results for four new fields (and reanalysis of two older fields) in the vicinity of the GSS.   We identify multiple kinematically cold features in their velocity distributions that we associate with the GSS or other photometric overdensities in M31's stellar halo.   These include the innermost unambiguous kinematical detection of the GSS and the first kinematical detection of GSS ``envelope'' debris.  We also present the kinematical detection of substructure in the region of the photometrically identified 60~kpc minor-axis arc \citep{ibata2007}, which some models of the collision suggest could be related to the GSS \citep{fardal2008}.  In addition, we further characterize the properties of a secondary kinematically cold component discovered by \citet{kalirai2006gss} in their study of a GSS core field at \rproj\,$=21$~kpc.  The kinematical and chemical properties of stars in the GSS presented here will provide improved observational constraints on the orbit and internal structure of the dwarf satellite progenitor and M31's potential.

The observations, reduction procedures, and our technique for identifying a clean sample of M31 RGB stars are 
described in \S\,\ref{sec:data}.  The properties of stars in individual spectroscopic fields are discussed in \S\,\ref{sec:ind_fields}: the kinematical components in each field are identified and their properties (mean velocity and velocity dispersion) are measured.   The properties of the GSS and other debris features are discussed and compared in \S\,\ref{sec:prop}.  Finally, the observations are discussed in the broader context of stellar halos in \S\,\ref{sec:discussion} and the main results of the paper are summarized in \S\,\ref{sec:gss_concl}. 
  
\section{Data}\label{sec:data}
In this section, we describe the data reduction and analysis techniques that 
produce the final M31 RGB sample analyzed in the following sections.  We 
describe the selection of fields from our spectroscopic 
survey (\S\,\ref{sec:fields}), give an overview of our photometric and 
spectroscopic observations and reductions (\S\S\,\ref{sec:obs}\,--\,\ref{sec:prev_pub}), and 
discuss the [Fe/H] measurements used in our survey (\S\,\ref{sec:gss_data_met}).
Finally, we describe the technique used to isolate the general M31 field 
population of RGB stars from 
MW dwarf stars and M31 dwarf spheroidal members (\S\,\ref{sec:gss_method}), and 
present the resulting velocity distributions of stars (\S\,\ref{sec:vels}).

\subsection{Field Selection}\label{sec:fields}

\begin{figure}[tb]
\epsscale{1.2}
\plotone{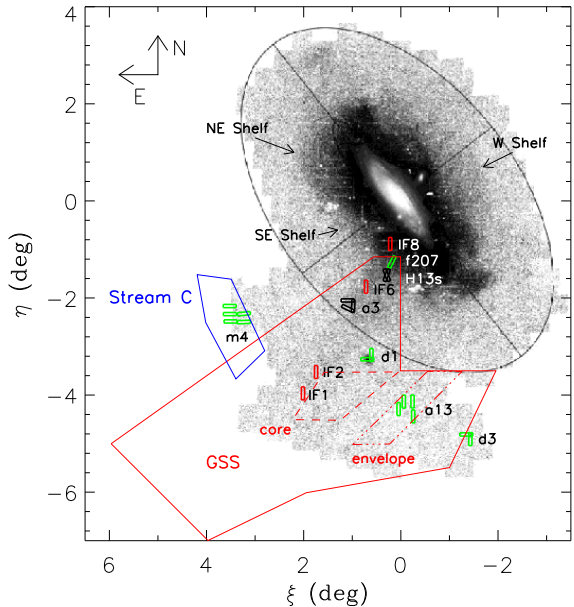}
\caption{
Locations (in M31-centric coordinates) of the Keck/DEIMOS spectroscopic fields discussed in this paper (\S\,\ref{sec:data}), 
overlaid on a star count map of M31 \citep{ibata2005}.  The GSS is the overdensity of
stars extending to the south-southeast of M31's center.  The approximate size and position 
angle of each DEIMOS spectroscopic slitmask is indicated by a rectangle.  
Black rectangles denote the location of masks from our survey of M31 for which velocity distributions have been 
previously published \citep[a3 and H13s;][]{guhathakurta2006,kalirai2006gss}, 
while green rectangles denote masks whose velocity distributions are being 
presented for the first time.
Red rectangles denote masks IF1, IF2, IF6 and IF8 from \citet{ibata2004}.   
The polygons show the regions chosen by \citet{ibata2007} for analysis of the photometric properties
of the GSS (red solid outline), GSS core (red dashed outline), GSS envelope (red dashed-dotted outline), and the 60~kpc southeast minor-axis arc ``stream C'' (blue solid outline).  
}
\label{fig:gss_roadmap}
\end{figure}

The data presented here represent a small subset of the full data set collected as part of our SPLASH Keck/DEIMOS spectroscopic survey of M31 RGB stars.  The subset was
selected based on location of the fields with respect to M31's GSS and other debris that may be related to the GSS (Fig.~\ref{fig:gss_roadmap}).  In relatively high-density regions of M31, ``fields'' consist of one spectroscopic mask (field f207) or two overlapping spectroscopic masks (field H13s), while in lower density regions fields consist of multiple masks designed from a single KPNO~4~m/Mosaic pointing (e.g., fields a13 and m4).  Fields d1 and d3 were initially targeted to probe the stellar kinematics of the dwarf spheroidals And~I and And~III (Kalirai et al., in preparation).  We are using them here to study the stellar kinematics of the underlying M31 field halo population, including possible GSS stars.

Fields f207, H13s, and a3 are located on/near the highest-surface brightness region of the GSS core (near its eastern edge) at \rproj\,$\sim 17$, 21, and 33~kpc, respectively.  Fields d1 and a13 are located to the west of the highest-surface brightness region of the GSS at \rproj\,$\sim 45$ and 58~kpc, respectively, but are still well within the broader region spanned by the GSS \citep{ibata2007}, 
while field d3, at \rproj\,$\sim 69$~kpc, is on the western edge of this region.  Finally, field m4 is located on M31's southeast minor axis at \rproj\,$\sim 57$~kpc, overlapping one of the minor-axis arc features (``stream C'') identified as stellar streams in the photometric survey of \citet{ibata2007}.  The latest models show that for certain collision geometries, similar arc morphologies on the minor-axis can be produced, and thus stream C could be related to the GSS \citep{fardal2008}.    

\subsection{Photometric and Spectroscopic Observations}\label{sec:obs}
Photometric catalogs for fields f207 and H13s were derived from images taken with 
the MegaCam instrument\footnote{MegaPrime/MegaCam is
a joint project of CFHT and CEA/DAPNIA, at the Canada-France-Hawaii Telescope
which is operated by the National Research Council of Canada, the Institut
National des Science de l'Univers of the Centre National de la Recherche
Scientifique of France, and the University of Hawaii.} 
on the CFHT 3.6-m telescope \citep[for details, see][]{kalirai2006gss}.  Images were obtained with the $g'$ and $i'$ filters.  The
observed stellar magnitudes were transformed to Johnson-Cousins $V$ and $I$ magnitudes
using observations of Landolt photometric standard stars \citep{kalirai2006halo}.  
Photometry for the remainder of the fields 
(a3, a13, m4, d1, and d3) was 
derived from images taken with the Mosaic Camera on the KPNO 4-m 
telescope\footnote{Kitt Peak National 
Observatory of the National Optical Astronomy Observatory is operated by the
Association of Universities for Research in Astronomy, Inc., under cooperative
agreement with the National Science Foundation.}.
The KPNO images were obtained with the Washington system
$M$ and $T_2$ filters and the intermediate-width DDO51 filter \citep{ostheimer2003}.  The observed magnitudes were transformed
to $V$ and $I$ magnitudes using the transformation equations of \citet{majewski2000}.

\begin{deluxetable*}{lcrrrlcccc}
\tabletypesize{\scriptsize}
\tablecolumns{10}
\tablewidth{0pc}
\tablecaption{Details of Spectroscopic Observations and Basic Results.}
\tablehead{\multicolumn{1}{c}{Mask} & \multicolumn{1}{c}{Projected}          & \multicolumn{2}{c}{Pointing center:} &
  \multicolumn{1}{c}{PA} & \multicolumn{1}{c}{Date of} & \multicolumn{1}{c}{Photometric} & 
  \multicolumn{1}{c}{\#\ Sci.}  & \multicolumn{1}{c}{\# } & \multicolumn{1}{c}{\#\ of M31} \\ 
       & \multicolumn{1}{c}{Radius}     & \multicolumn{1}{c}{$\alpha_{\rm J2000}$} &
\multicolumn{1}{c}{$\delta_{\rm J2000}$} & \multicolumn{1}{c}{($^\circ$E of N)}
& \multicolumn{1}{c}{Obs. (UT)} & \multicolumn{1}{c}{Bandpasses\tablenotemark{a}} & \multicolumn{1}{c}{targets\tablenotemark{b}}  &
 \multicolumn{1}{c}{Stars\tablenotemark{b,c}} &
\multicolumn{1}{c}{Stars\tablenotemark{b,d}} \\
     & \multicolumn{1}{c}{(kpc)} & \multicolumn{1}{c}{($\rm^h$:$\rm^m$:$\rm^s$)} &
\multicolumn{1}{c}{($^\circ$:$'$:$''$)} &  & & & & & }
\startdata
f207\_1 & 17.4 & 00:43:42.64 & +40:00:31.6 & $-27.0$ & 2005 Nov 4 & $g'$, $i'$ & 202 & 123 & 114\\
H13s\_1\tablenotemark{e} & 21.3 & 00:44:14.76 & +39:44:18.2 & +21.0 & 2004 Sep 20 & $g'$, $i'$ & 134 & 100 &  87\\ 
H13s\_2\tablenotemark{e} & 21.3 & 00:44:14.76 & +39:44:18.2 & $-21.0$ & 2004 Sep 20 & $g'$, $i'$ & 138 & 121 & 98\\ 
a3\_1\tablenotemark{f} & 33.9  & 00:48:21.16 & +39:02:39.2 & +64.2 & 2002 Aug 16 & $M$, $T_2$, DDO51 & 85 & 20 & 17 \\ 
a3\_2\tablenotemark{f} & 32.5 & 00:47:47.24 & +39:05:56.3 & +178.2 & 2002 Oct 11 & $M$, $T_2$, DDO51 & 81 & 32 & 28\\ 
a3\_3\tablenotemark{f} & 31.9 & 00:48:23.17 & +39:12:38.5 & +270.0 & 2003 Oct 26 & $M$, $T_2$, DDO51 & 83 & 43 & 30 \\ 
d1\_1 & 44.1 & 00:45:48.61 & +38:05:46.8 & 0.0 & 2005 Nov 5 & $M$, $T_2$, DDO51 & 155 & 91 & 21 \\
d1\_2 & 45.6 & 00:46:13.95 & +38:00:27.0 & $+90.0$ & 2006 Sep 16 & $M$, $T_2$, DDO51 & 149 & 96 & 28 \\
a13\_1 & 58.5 &  00:42:58.34 & +36:59:19.3 & 0.0 & 2003 Sep 30 & $M$, $T_2$, DDO51 & 80 & 25 & 10\\  
a13\_2 & 60.6 & 00:41:28.27 & +36:50:19.2 & 0.0 & 2003 Sep 30 & $M$, $T_2$, DDO51 & 71& 23 & 10\\
a13\_3 & 56.4 &  00:42:25.76 & +37:08:25.4 & 0.0 & 2005 Nov 5 & $M$, $T_2$, DDO51 & 78 & 29 & 13\\
a13\_4 & 56.5 & 00:41:32.15 & +37:08:37.4 & 0.0 & 2005 Nov 5 & $M$, $T_2$, DDO51 & 86 & 32 & 11\\
d3\_1 & 68.3 & 00:36:03.83  & +36:27:27.4 & $+90.0$ & 2005 Sep 8 & $M$, $T_2$, DDO51 & 120 & 89 & 5 \\
d3\_2 & 69.8 & 00:35:39.61 & +36:21:41.8 & 0.0 & 2005 Sep 8 & $M$, $T_2$, DDO51 & 122 & 84 & 8 \\
m4\_1 & 59.3 & 01:00:40.24 & +38:42:07.6 & $-90.0$ & 2006 Sep 20 & $M$, $T_2$, DDO51 & 72 & 30 & 13\\
m4\_2 & 58.3 & 01:00:46.90 & +38:51:05.9 & $-90.0$ & 2006 Sep 21 & $M$, $T_2$, DDO51  & 81 & 22 & 5\\
m4\_3 & 57.1 & 01:00:50.72 & +39:01:00.6 & $-90.0$ & 2006 Nov 20 & $M$, $T_2$, DDO51 & 90 & 40 & 15\\
m4\_4 & 56.3 & 00:59:15.92 & +38:42:12.0 & $-88.3$ & 2006 Nov 20 & $M$, $T_2$, DDO51 & 70 & 31 & 6\\
m4\_5 & 54.9 & 00:59:18.67 & +38:52:22.5 & $-84.0$ & 2006 Nov 23 & $M$, $T_2$, DDO51 & 75 & 38 & 14\\
\enddata
\tablenotetext{a}{The catalogs used for spectroscopic target selection were derived from photometry obtained in the listed photometric bandpasses (\S\,\ref{sec:obs}). }
\tablenotetext{b}{A number of targets were observed on two different masks.
The total number of {\it unique} science targets/stars/M31 RGB stars in fields H13s, a3, d1, and d3 
is therefore less than the reported number.  The number of stars with velocities recovered in two separate observations is 9 in H13s, 7 in a3, 3 in d1, and 11 in d3.  Of these, 7 stars
in H13s, 6 in a3, and 1 in d1 are M31 RGB field stars.  Of the duplicates in the dwarf satellite fields, 1 each in d1 and d3 are MW dwarf stars; the remainder are dSph members.}
\tablenotetext{c}{The number of stellar spectra with secure velocity measurements.}  
\tablenotetext{d}{The number of M31 RGB stars is defined as the number of stars
that are identified as secure or marginal M31 RGB stars by the \citet{gilbert2006}
diagnostic method (\S\,\ref{sec:gss_method}).  For the d1 and d3 masks, additional selection criteria have been applied to exclude members of the dSphs (\S\,\ref{sec:and_fields}).}
\tablenotetext{e}{Previously published in \citet{kalirai2006gss}, the data presented here have undergone new reductions with our improved pipeline (\S\,\ref{sec:prev_pub}).  The sample of M31 RGB stars presented here is 70\% larger than in \citet{kalirai2006gss}.}
\tablenotetext{f}{Previously published in \citet{guhathakurta2006}, the data presented here have undergone new reductions with our improved pipeline  (\S\,\ref{sec:prev_pub}).  The data was originally published without the benefit of the \citet{gilbert2006} diagnostic method for determining M31 membership (\S\,\ref{sec:gss_method}).}
\label{table:gss_masks}
\end{deluxetable*}
\def\arraystretch{1.0}

 

The DDO51 filter measures the surface-gravity sensitive Mgb and MgH stellar absorption features.  Therefore photometry in the intermediate-width DDO51 band, when combined with photometry in the $M$ and $T_2$ bands, can be used as a discriminant of stellar surface-gravity.  This allows
photometric pre-selection of spectroscopic targets that are likely to 
be red giant branch (RGB) stars in M31 (as opposed to foreground MW dwarf stars).  For masks designed from Mosaic photometry, the spectroscopic target selection algorithm gave highest priority to objects with high photometric RGB probability [as determined from the
star's position in the ($M-$\,DDO51) versus ($M-T_2$) color-color diagram \citep[e.g.,][]{palma2003,majewski2005}], star-like morphology (based on the DAOPHOT parameters {\tt chi} and {\tt sharp}), and with $I_0$ magnitudes within $\sim 1$~mag of the tip of M31's RGB.  The available space remaining on the mask was then packed with lower priority ``filler'' targets.  All spectroscopic targets were restricted to the brightness range $20<I_0<22.5$.  In fields imaged with CFHT/MegaCam, spectroscopic targets
were selected based on their $I_0$ magnitude and the SExtractor morphological 
criterion {\tt stellarity}.  Details of the spectroscopic target selection procedure for fields with and without DDO51 photometry are given by \citet{guhathakurta2006} and \citet{kalirai2006gss}, respectively.

Spectra were obtained using the DEIMOS spectrograph on the Keck~II 10-m telescope,
using the 1200~line~mm$^{-1}$ grating, which has a dispersion of 
$\rm0.33~\AA$~pix$^{-1}$.  The observations were obtained over five observing seasons (Fall 2002\,--\,2006) and are summarized 
in Table~\ref{table:gss_masks}.  The spectra were reduced using modified versions of the
{\tt spec2d} and {\tt spec1d} software developed at the University
of California, Berkeley in support of the DEEP2 galaxy redshift survey\footnote{{See \tt
http://astron.berkeley.edu/$\sim$cooper/deep/spec2d/primer.html},
\newline\indent
{\tt http://astron.berkeley.edu/$\sim$cooper/deep/spec1d/primer.html}
}.  These routines perform standard spectral reductions: flat-fielding, night-sky emission line removal,
extraction of the two-dimensional spectra, and cross-correlation of the one-dimensional spectra with template
stellar spectra.  The template library includes stellar spectra obtained with DEIMOS and galaxy templates from the Sloan Digital Sky Survey.  The cross-correlation procedure has 
been tested extensively against radial velocity standards \citep{simon2007}, 
and successfully reproduces known velocities.  All spectra and cross-correlations 
were visually inspected, and only spectra with secure velocity measurements 
were included in further analysis.  The observed velocity was corrected 
both for the motion of the earth around the sun (heliocentric correction) 
and for imperfect centering of the star within the slit 
\citep[based on the observed position of the atmospheric A-band absorption feature relative to night sky emission lines;][]{simon2007,sohn2007}. 
The typical error in the velocity measurements are $\sim 6$~\kms\ for the M31 RGB stars presented here.  It is calculated by adding in quadrature the random velocity measurement error (estimated from the cross-correlation routine) and a systematic error of 2.2~\kms\ \citep[estimated via repeat observations of stars;][]{simon2007}.

\subsection{Previously Published Fields}\label{sec:prev_pub}
The fields H13s and a3 have been previously analyzed by \citet{kalirai2006gss} 
and \citet{guhathakurta2006}, respectively.  These were among the first 
fields observed and reduced in our M31 spectroscopic survey.  Since the original reduction of 
these fields, we have optimized our pipeline for reducing and measuring the velocities 
of stellar objects.  It now includes a greater number of spectral templates (obtained with the same
instrument as the science spectra) for measuring velocities via cross-correlation
as well as a correction for imperfect centering of the star in the slit (\S\,\ref{sec:obs}).  
These improved data reduction 
and cross-correlation routines result in smaller systematic and random velocity errors.  Velocities obtained with the improved cross-correlation procedure are shifted by $\sim +20$~\kms\ from the velocities obtained with the original cross-correlation pipeline.  The typical measured velocity error of M31 RGB stars in these fields has been reduced from $\sim 15$~\kms\ to $\sim 6$~\kms.

The data in fields a3 and H13s presented here have
been re-reduced using the new cross-correlation routine.  All papers published 
by our group from 2007 onwards utilize our improved reduction pipeline.  
We have also recovered a significant 
fraction of spectra in field H13s that had failed in our original 
reduction pipeline. The sample of M31 RGB stars in field H13s is now 70\% larger 
than the sample presented by \citet{kalirai2006gss}.  

In addition to the improvement in the reduction pipeline, the data in field a3 have benefited from an improved method to separate M31 RGB stars from foreground MW dwarf stars \citep[][see \S\,\ref{sec:gss_method} below]{gilbert2006}.  This method had not yet been fully developed when these data were first presented. It is particularly important for securely identifying RGB stars in the kinematically broad spheroid of M31, whose radial velocities can overlap the tail of the MW dwarf star velocity distribution.

\subsection{Photometric and Spectroscopic Metallicity Estimates}\label{sec:gss_data_met}

We estimate metallicities using both photometric and spectroscopic 
techniques for each star in our M31 survey.  Like \citet{kalirai2006halo}, 
we estimate photometric metallicities, \fehp, by comparing the
star's position in the \ivio\ color-magnitude diagram (CMD) to a finely 
spaced grid of theoretical
isochrones [$t=12$~Gyr, [$\alpha$/Fe]=0.0; \citet{vandenberg2006}].  The
observed stellar magnitudes are transformed into Johnson-Cousins $V$ and $I$ 
magnitudes using the transformations described in \S\,\ref{sec:obs}, and 
extinction/reddening corrections based on the \citet{schlegel1998} dust maps 
are applied on a star-by-star basis. 
A distance modulus of $(m-M)_0=24.47$ \citep[corresponding to a distance 
to M31 of 783~kpc;][]{stanek1998} is applied to the isochrones.  
Although several systematic differences in our photometric metallicity 
estimates could exist between the data from the two different filter systems 
\citep{kalirai2006halo}, tests of the data show that any such differences are small 
compared to other sources of systematic error in our photometric metallicity 
estimates (discussed below).  

The spectroscopic metallicity estimates, \fehs, are based on the strength 
of the \caii-triplet absorption lines at $\lambda\sim 8500$\AA\, using 
the empirical calibration relations of 
\citet{rutledge1997pasp1,rutledge1997pasp2} and assuming a horizontal 
branch magnitude of $I_0=25.17$ for M31 stars.  The calibration is derived 
from RGB stars in MW globular clusters that span a range of metallicities.  

It is reassuring that in general our photometric metallicity estimates are in good
agreement with the spectroscopic metallicity estimates based on the 
strength of the \caii\ absorption line triplet.  The
individual \caii\ EW measurements and derived \fehs\ estimates have large random errors
\citep{gilbert2006} due to the low typical signal-to-noise of our M31 RGB
spectra ($S/N\sim 15$~per \AA).  Since the \fehs\ estimates are very noisy,
we can only make a meaningful comparison if we bin our sample of stars in terms
of \fehp\ and compute the average \fehs\ in each bin \citep[Fig.\ 7 of][]{kalirai2006halo}.  
We thus use photometric rather
than spectroscopic metallicity estimates for the analysis presented here.

Although the random error is smaller for \fehp\ than \fehs, there are 
substantial systematic errors in \fehp\ 
due to differences between the age and alpha-enrichment of a star and the 
adopted isochrones.  Moreover, there are substantial variations between 
the isochrones computed by different groups. 
Photometric metallicity estimates based on the Padova \citep{girardi2002} 
and Yale-Yonsei \citep[Y$^2$;][]{demarque2004} isochrone sets vary from 
those based on the Victoria-Regina \citep[VR;][]{vandenberg2006} isochrones 
by a median amount of  $\sim 0.1$~dex.  The variation across the isochrone 
sets is largest for the bluest stars ($\sim 0.3$\,--\,0.5~dex).   

There are no observational constraints on the alpha-enrichment of M31 
halo stars, but 
the assumption of a single age for the stellar population is known to be 
incorrect.  Deep HST/ACS observations of fields in M31's stellar halo and 
the GSS that reach below the main-sequence turnoff have shown that the 
age of the stellar population ranges from $\sim 8$\,--\,13~Gyr in the halo 
and $\sim 6$\,--\,13~Gyr in the GSS; $\sim 33$\% of the population is 
younger than 10~Gyr in M31's halo, while $\sim 70$\% of the population is 
younger than 10~Gyr in the GSS field \citep{brown2006apj,brown2007}.   
These studies find that in general the intermediate-age stars are relatively 
metal-rich ($-0.75\lesssim$\,[Fe/H]\,$\lesssim 0.4$) while the older 
populations are relatively metal-poor ([Fe/H]\,$\lesssim -1.0$).  

Our photometric metallicity estimates are therefore likely to be biased 
low for the most metal-rich stars by our assumption of an old (12 Gyr), 
coeval stellar population, since the HST studies show that these metal-rich 
stars are typically of intermediate-age.  Changing the isochrone age from 
12 to 6 Gyr results in a $\sim +0.15$~dex shift in the derived stellar 
metallicity for stars with \fehp\,$ > -1.0$, and a $\sim +0.3$~dex shift 
for stars with $-2.3 < $\,\fehp\,$<-1.5$.  Although the systematic bias 
in the derived metallicity estimates introduced by assuming an old 
stellar population is large for intermediate-age, metal-poor stars, 
the HST studies show that such stars are rare in both M31's halo and the 
GSS \citep{brown2006apj}.

\subsection{Isolation of the Field M31 RGB Sample}\label{sec:gss_method}
Even in fields with DDO51-based photometric preselection of likely M31 
RGB stars (\S\,\ref{sec:obs}), MW dwarf stars along the line of sight to M31 
still constitute a significant fraction of the recovered stellar spectra.  
Since the radial velocity distributions of MW dwarf stars and M31 RGB stars
overlap, we use a set of photometric and spectroscopic diagnostics to 
determine the probability that an individual star is a red giant at the 
distance of M31 or a dwarf star in the MW.  The 
diagnostics are (1) radial velocity, (2) photometry in the Washington $M$ 
and $T_2$ bands and the DDO51 band, (3) \nai\ EW versus \vio\ color,
(4) position in an \ivio\ CMD and (5) spectroscopic versus photometric \feh\ 
estimates.  This technique for separating the M31 RGB and MW dwarf star 
populations is presented and tested in detail by \citet{gilbert2006}.  
We briefly summarize it here. 

Each of the five diagnostics provides separation between M31 RGB stars 
and MW dwarf stars based on different physical parameters.  The combination 
of the five diagnostics provides a powerful test of whether an individual 
star is an M31 red giant or MW dwarf.  While M31 RGB stars have a broad 
velocity distribution around M31's systemic velocity of 
$v_{\rm sys}=-300$~\kms, MW dwarf stars tend to share the sun's rotation 
about the MW's center and have $v_{\rm helio}\approx 0$~\kms.  As discussed 
in \S\,\ref{sec:obs}, the combination of the $M$, $T_2$, and DDO51 photometry 
is sensitive to the difference in surface gravity between RGB and dwarf stars.  
The strength of the \nai\ doublet at $\lambda\sim 8190$\AA\ is both 
surface-gravity and temperature sensitive.  It provides good separation of 
RGB and dwarf stars with \vio\,$\gtrsim 2$.  M31 RGB stars follow the 
RGB isochrones in the CMD, while MW dwarf stars are more broadly distributed 
in CMD space.  Finally, due to the fact that both the \fehp\ and \fehs\ 
estimates are based on the assumption that the star is an M31 RGB star, 
these estimates agree only if the star is in fact an RGB star at the 
distance of M31.

\begin{figure}[tb]
\epsscale{1.0}
\plotone{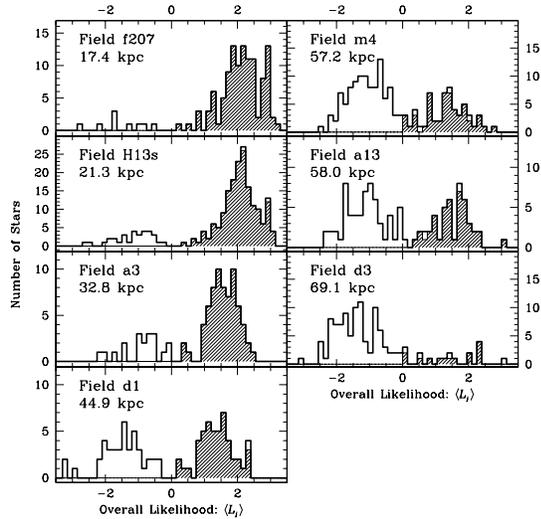}
\caption{
Overall likelihood distributions for stars in each of the fields analyzed in this paper (\S\,\ref{sec:gss_method}).  The shaded histograms denote stars identified as M31 RGB stars.  Stars with \olike\,$>0$ that are not shaded are stars that, despite having physical characteristics that are consistent with the M31 RGB population, have been removed from the M31 RGB sample due to their very blue \vio\ colors.  In the d1 and d3 fields, only stars {\it not} associated with the dSph galaxies are included in this distribution (\S\,\ref{sec:and_fields}). 
}
\label{fig:likelihoods}
\end{figure}

A star $i$ is given a likelihood in a given diagnostic $j$ of being an 
M31 RGB star, defined as $L_{ij}={\rm log}(P_{\rm giant}/P_{\rm dwarf})$, 
where $P$ is the probability the star is an RGB or dwarf star based on a 
comparison of the star's position in that diagnostic space to probability 
distribution functions based on empirical training sets of definite M31 
RGB and MW dwarf stars \citep[][\S\,3]{gilbert2006}.
The individual likelihoods are combined to form an overall likelihood \olike. If 
\olike\,$>0$, it is more probable that the star is an RGB star than a dwarf 
star, and vice versa for \olike\,$<0$.  Figure~\ref{fig:likelihoods} 
shows the overall likelihood distributions of stars in each of our fields.
For the purposes of the current analysis, we use stars with \olike\,$>0$--- 
i.e.,  more likely to be an M31 RGB star than an MW dwarf star.  Furthermore, 
we include only stars whose positions in the \ivio\ CMD are within the 
isochrone grid (\fehp\,$=-2.31$ to 0.0) to within the average photometric 
error of the sample\footnote{In the terminology of \citet{gilbert2006}, we 
include class +1, +2, and +3 stars in the following analysis, and 
exclude class $-2$ stars from our M31 RGB sample.}.

It is possible that stars with \olike\,$>0$ that are significantly bluer 
than the isochrone grid in the CMD may be asymptotic 
giant branch stars at the distance of M31 or M31 RGB stars with flawed 
photometry.  However, MW dwarf stars along the line of sight with large 
negative velocities ($v_{\rm los}\lesssim -150$) tend to have blue \vio\ 
colors, as these stars are generally main-sequence turn-off stars in the 
MW's halo. Since several of the diagnostics have limited differentiating 
power between RGB and dwarf stars for very blue stars 
[\vio\,$\lesssim-1.2$], the \olike\ values of these stars are strongly 
influenced by their M31 RGB-like velocities \citep[][\S\,4.1.2]{gilbert2006}.  
We therefore choose to be conservative and exclude these very blue stars 
from our M31 RGB sample.  These stars represent only 2\% of the total 
stellar sample.

In \citet{gilbert2007}, we used a large sample of RGB stars in M31's 
inner spheroid (\rproj\,$\lesssim 30$~kpc) to investigate the effect of
removing the velocity diagnostic from our selection of M31 RGB and MW dwarf
stars.  This sample was ideal for this analysis since the density of 
RGB stars in M31's inner regions is much higher than the density of 
MW stars along the line-of-sight.  We found that while the exclusion of 
the velocity diagnostic
does increase the completeness of the M31 RGB sample at velocities between
$-100$ and 0~\kms, it also increases the level of MW dwarf 
star contamination at these velocities.  Some of the fields presented
here are at large distances from M31's center (\rproj\,$\sim 60$~kpc), where
the density of MW dwarf stars along the line-of-sight is significantly higher
than the density of M31 RGB stars.  We thus choose lower contamination
over higher completeness for the present work, and include the velocity 
diagnostic in our selection of M31 RGB stars.  This affects only 
stars with velocities in the range of the MW dwarf distribution.  Since
we do not fit the velocity distribution of M31's spheroid in this work 
and the kinematically cold substructures presented below have mean velocities 
$\lesssim -250$~\kms\ (\S\,\ref{sec:ind_fields}), incompleteness of the M31 RGB 
sample at velocities near zero will only affect the estimated fraction of the stars in 
M31's kinematically broad spheroid.   
Furthermore, this effect will be small since the portion
of velocity-space where we expect incompleteness due to the use of the 
velocity diagnostic is both in the tail of the M31 spheroid velocity 
distribution and is small compared to the full velocity space used 
in the fits to the velocity distributions in \S\,\ref{sec:ind_fields}. 

\subsubsection{Andromeda Satellite Fields}\label{sec:and_fields}

The dSph And~I is located within the area spanned by the GSS, 
while the dSph And~III is located just outside this region 
(Fig.~\ref{fig:gss_roadmap}, marked by the ``d1'' and ``d3'' spectroscopic masks, respectively).  In addition to dSph RGB stars and MW dwarf 
stars, M31 ``field'' stars (RGB stars that are physically associated with 
the halo of M31 rather than with the dSph) are expected to be present in 
our spectroscopic sample.  The half-light radii of the dSphs And~I and And~III 
are 2.8$'$ (600~pc) and 1.6$'$ (360~pc), respectively \citep{mcconnachie2006}, 
therefore much of the area of each 4$'$\,$\times$\,16$'$ DEIMOS slitmask 
primarily targets the M31 field star population.

\begin{figure}[tb]
\epsscale{1.0}
\plotone{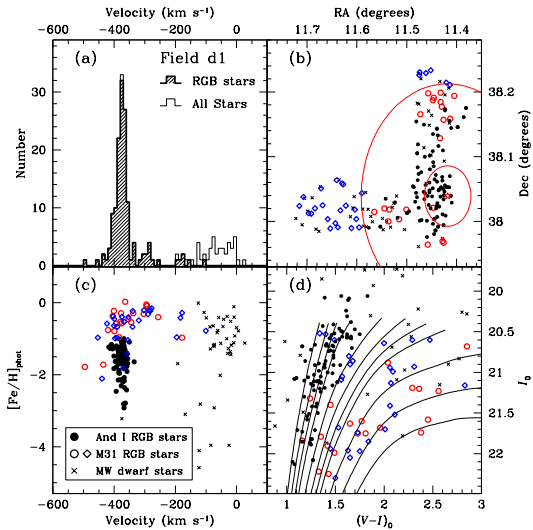}
\caption{M31 field star selection in the d1 spectroscopic field (\S\,\ref{sec:and_fields}).  
({\it a}) 
Line-of-sight velocity distribution of all stars 
(thin open histogram), and stars identified as RGB stars by the likelihood method
described in \S\,\ref{sec:gss_method} (thick, shaded histogram).  In addition to the prominent kinematically-cold peak formed by dSph members, there are RGB stars present in this field with a wide range of velocities, as expected for stars associated with M31's spheroid.   
({\it b}) 
Sky positions of the RGB stars in field d1. The center (red star), half-light radius (inner ellipse) and tidal radius (outer ellipse) of the And~I dSph are marked \citep{mcconnachie2006}.  RGB stars at or beyond the tidal radius of the dSph are designated as M31 field stars (open diamonds).  Open circles denote red giants designated as M31 field stars by virtue of their separation from the locus of dSph members in the $v$--\fehp\ plane (panel {\it c}).  Filled circles denote RGB stars that do not meet our selection criteria for M31 field stars.  The vast majority of these stars are dSph members. Crosses denote MW dwarf stars.
({\it c}) 
Velocity vs.\ \fehp\ for all stars.  
RGB stars associated with the dSph form 
a tight locus in this space.  A second population 
of RGB stars with higher \fehp\ values (\fehp$>-1.0$) and a wider spread in 
velocities, most of which are beyond the tidal radius of the dSph, are more likely associated with M31's spheroid.  
({\it d}) 
Location of M31 field stars and dSph members in the \ivio\ CMD.  Theoretical
isochrones shifted to the distance of M31 have been overlaid on the data.  The isochrones are for an age of 12~Gyr, $[\alpha/\rm Fe]=0$, and span a range of [Fe/H] from $-2.31$ to $0.0$ \citep{vandenberg2006}. 
}
\label{fig:and1}
\end{figure}

\begin{figure}[tb]
\epsscale{1.0}
\plotone{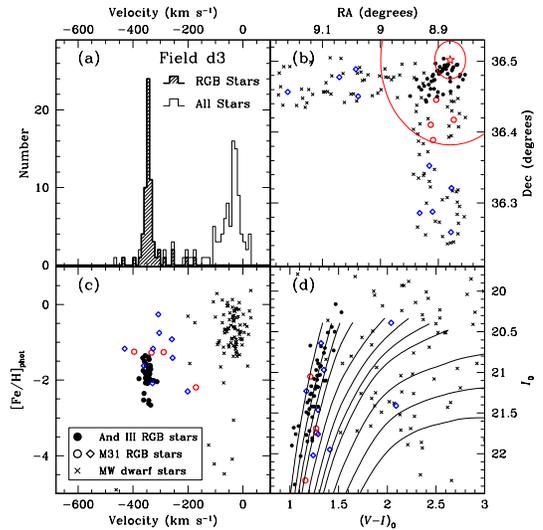}
\caption{Same as Figure~\ref{fig:and1}, for field d3 located on the dSph And~III.
}
\label{fig:and3}
\end{figure}

M31 field RGB stars are expected to have different velocity, metallicity, and 
spatial distributions than the dSph members.  Figures~\ref{fig:and1}({\it a}) and \ref{fig:and3}({\it a}) display the velocity distributions of all stars observed in the d1 and d3 fields, respectively (thin open histograms).  The bold shaded histograms denote the stars identified as red giants using the method described in \S\,\ref{sec:gss_method}.    
Although the majority of the RGB stars in the two dSph fields are part of the prominent kinematically-cold velocity distributions of the And~I and And~III dSphs (at $v\sim -375$ and $-345$~\kms, respectively), there are also RGB stars spanning a wide range of velocities in each field, as expected for stars that are associated with M31's spheroid. 

We utilize the spatial, kinematic, and chemical properties of the RGB stars to  identify individual RGB stars as likely M31 field stars.  The spatial distributions of stars observed in the d1 and d3 fields are displayed in Figures~\ref{fig:and1}({\it b}) and \ref{fig:and3}({\it b}).  The center of each dSph is marked by an open star.  The inner and outer ellipses denote the half-light radii and tidal radii, respectively, of each dSph \citep{mcconnachie2006}.  All RGB stars beyond the tidal radius of the dSph are designated as M31 field stars (open diamonds).  It is possible that some of these stars were tidally stripped from the dSph, in which case they are recent additions to M31's spheroid.  

Figures~\ref{fig:and1}({\it c--d}) and \ref{fig:and3}({\it c--d}) demonstrate that stars associated with the dSph galaxies (filled circles) form fairly tight loci in the $v$--\fehp\ plane and the \ivio\ CMD.  A small fraction of the RGB stars within the tidal radii of the dSphs are designated as likely M31 RGB stars by virtue of having velocities and/or \fehp\ estimates that are well removed from the dSph locus (open circles).  All but one of these stars are beyond the half-light radius of the dSph [Figs.~\ref{fig:and1}({\it b}) and \ref{fig:and3}({\it b})].  In the d3 field, these stars are removed from the mean velocity of the dSph by more than 8 times the measured width of the dSph velocity distribution ($\sigma_v$).  The RGB star in the d1 field at $v=-438$~\kms\ that is within the tidal radius of the dSph is $5.8\sigma_v$ removed from the mean velocity of And~I.  The full RGB population in the d1 field appears to be comprised primarily of two distinct populations: a relatively metal-poor (\fehp$<-1.0$) population spanning a narrow velocity range (the And~I dSph), and a relatively metal-rich (\fehp$>-1.0$) population with a broad range of velocities.   The stars beyond the tidal radius of And~I comprise the majority of the more metal-rich population [Fig.~\ref{fig:and1}({\it c--d})], indicating that this population is not associated with the dSph.  We therefore designate all RGB stars in d1 with \fehp$>-1.0$ as likely M31 field stars.

The selection criteria for M31 field stars described above yields an incomplete sample, as M31 field stars that fall both within the tidal radius of the dSph and near the region of $v$--\fehp\ space occupied by the dSph members are not included in our sample.  However, the fraction of M31 field stars that satisfy both these criteria is expected to be small, since (i) dSph members span a very narrow range in velocity and (ii) the surface brightness of each dSph becomes increasingly dominant over M31's spheroid as one moves towards the center of the dSphs, resulting in an increasingly greater likelihood that spectroscopic slits are placed on a dSph member rather than an M31 field star.  Only the M31 field stars will be analyzed in the current work.  The properties of the dSph members in these fields will be discussed in a future paper (Kalirai et al., in preparation).

\subsection{Velocity Distributions}\label{sec:vels}

\begin{figure}[tb]
\epsscale{1.0}
\plotone{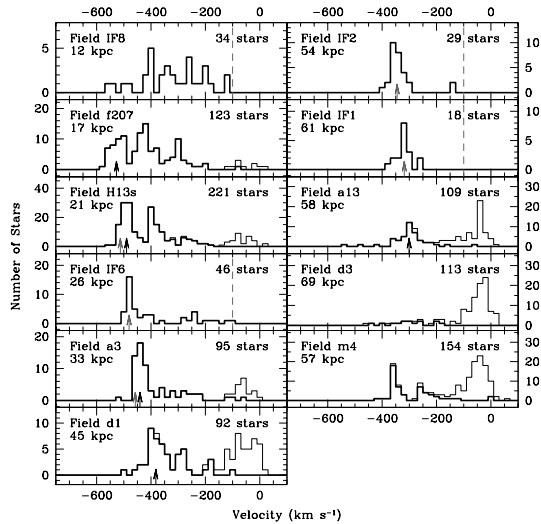}
\caption{Line-of-sight, heliocentric velocity distributions for stars in each of the 
fields shown in Figure~\ref{fig:gss_roadmap}, displayed in order of increasing distance
along the giant southern stream (GSS).  The velocity distributions of stars 
identified as M31 RGB stars (\S\,\ref{sec:gss_method}) are shown
by the bold histograms.  The velocity distributions of all observed stars (including stars identified as MW dwarfs) are 
shown as thin histograms.  Stars associated with the dSphs 
And~I and And~III have been removed from the velocity distributions.
Arrows mark the mean velocity of the GSS: 
black arrows denote velocities measured in this work (\S\,\ref{sec:ind_fields}), 
while grey arrows denote 
previously published velocities.  The mean velocity of the GSS becomes 
increasingly negative as the stream approaches the center of M31.  In addition to the
prominent kinematical peaks identified as the GSS, several of our spectroscopic fields 
show evidence of other kinematical substructures (fields f207, H13s, and m4; \S\,\ref{sec:ind_fields}).  Each panel 
is labeled with the projected radial distance of the field from M31's center 
and the total number of stars with radial velocity measurements.  The majority 
of the data come from the SPLASH Keck/DEIMOS survey of M31's stellar halo (fields f207, 
H13s, a3, d1, a13, m4, and d3).    The velocities in fields IF8, 
IF6, IF2, and IF1 were obtained from 
\citet{ibata2004}; only stars with $v < -100$\kms\ are available from the literature (dashed lines).  M31's systemic velocity is $v_{\rm sys}=-300$~\kms.
}
\label{fig:gss_velhist}
\end{figure}


Figure~\ref{fig:gss_velhist} presents the line-of-sight heliocentric 
velocity distributions of all stars (M31 red giants and MW dwarfs) 
observed in our fields, as well as the velocity 
distributions of stars in the four GSS spectroscopic fields 
published by \citet{ibata2004} (fields IF1, IF2, IF6, and IF8).  In the 
innermost fields, M31 RGB stars greatly outnumber MW dwarf stars along the 
line-of-sight, while the majority of stars observed in the outermost fields are 
MW dwarf stars due to the decreasing density of M31's halo with \rproj.

Most of the fields show clear evidence of kinematically cold
features in their stellar velocity distributions in addition to stars
with a wide range of velocities (M31's kinematically broad spheroid).  
In fact, several fields appear to have multiple kinematically cold 
components (e.g., f207, H13s, and m4). 
The GSS is a prominent structure in the stellar velocity distributions 
of fields near the eastern edge of the high surface brightness 
GSS core (f207, H13s, IF6, a3, IF2, and IF1; Fig.~\ref{fig:gss_roadmap}).
The detection of the GSS in field f207 is the innermost secure 
kinematical detection of the GSS to date.  
There are also prominent kinematically cold peaks in fields d1 and a13 
with velocities in keeping with the expected velocity of the GSS. 
We will show below that these are also attributable to the GSS, and are 
the first kinematical detections of the GSS in fields to the west of its 
prominent, eastern edge (\S\,\ref{sec:offgss_west} \& \S\,\ref{sec:prop_core}).

\section{Identification of Substructure in Individual Fields}\label{sec:ind_fields}

This section discusses in detail the stellar velocity distributions 
and evidence for the presence of kinematically cold components in each of 
our Keck/DEIMOS spectroscopic fields (Fig.~\ref{fig:gss_velhist}).  
The velocity and metallicity distributions of M31 RGB stars 
are analyzed in order to identify substructure in each field, and 
maximum-likelihood fits to the velocity distributions are performed 
to characterize the kinematical properties of the substructure, including its
mean velocity, velocity dispersion, and the fraction of stars belonging to it.  
The fields near the eastern edge of the GSS are discussed first (f207, H13s,
and a3; \S\,\ref{sec:ongss}), followed by the fields to the west (d1, a13, 
and d3; \S\,\ref{sec:offgss_west}) and the field on the 60~kpc minor-axis 
arc (m4; \S\,\ref{sec:m4}).  A broader analysis of the kinematical and 
chemical properties of the identified substructures in relation to each other
and a discussion of the origins of the  
features will be presented in \S\,\ref{sec:prop}.

\subsection{Characterizing Kinematical Substructure}\label{sec:cks}
We use a combination of kinematics and metallicities to identify
the number of kinematically-cold substructures in each field.  We then fit 
a multiple-component Gaussian to the kinematical distribution using
a maximum-likelihood technique.  The maximum-likelihood multi-Gaussian 
fits in each field are discussed below and summarized in 
Table~\ref{table:gss_mlfits}.   Quoted errors are 90\% confidence limits from 
the maximum-likelihood analysis.  Although the true shapes of the velocity 
distributions of the various kinematical components in M31's spheroid 
are likely different from pure Gaussians, they provide 
a convenient means of characterizing the mean velocity and velocity
dispersions of kinematical features.  

We assume a kinematically broad distribution of velocities 
corresponding to M31's well-mixed spheroid is present in each field.  
To date, only two spectroscopic studies have measured the velocity dispersion of
M31's spheroid.  \citet{gilbert2007} used a sample of $\sim 1000$ M31 RGB
stars in spectroscopic fields from \rproj\,$=9$\,--\,30~kpc 
along M31's minor axis. They measured
a velocity dispersion of \sigvsph\,$=129$~\kms\ for
M31's spheroid, and saw no evidence for a 
change in the velocity dispersion with radius between 9\,--\,30~kpc.
The stars were selected using the
technique described in \S\,\ref{sec:gss_method}, but without use of the 
velocity diagnostic.  
\citet{chapman2006} measured an average velocity dispersion of 
\sigvsph\,$=126$~\kms\ 
for M31's spheroid, consistent with the \citet{gilbert2007} value.  
Their measurement was based on a sample of $\sim 800$ M31 stars selected
using windows in line-of-sight velocity space designed to remove foreground
Milky Way stars, M31 disk stars, and substructure.  
They also fit a model in which the velocity dispersion of M31's spheroid
decreases with increasing projected distance.  Using this model,  
they find that M31's spheroid is expected to have a velocity 
dispersion of  \sigvsph\,$=137$~\kms\ 
at \rproj$=17$~kpc, decreasing to  \sigvsph\,$=100$~\kms\ at \rproj$=60$~kpc.

Since the number of M31 spheroid stars at large radii ($\gtrsim 40$~kpc) 
in the \citep{chapman2006} study is small and a significant portion of 
velocity space was not used for measuring the change in the velocity dispersion 
with radius, we assume a constant value for the dispersion of M31's
spheroid in analyzing the velocity distribution of our fields 
(\mvsph\,$=-300$~\kms, \sigvsph\,$=129$~\kms).  
However, for the fields at larger radii (\rproj$>40$~kpc), we will 
comment on the effect on the fit of assuming a 
smaller velocity dispersion for M31's spheroid. 


\begin{deluxetable*}{lrrrrrrl}
\tabletypesize{\small}
\tablecolumns{8}
\tablewidth{0pc}
\tablecaption{Results of Maximum Likelihood Fits: Fields with One or More Kinematically Cold Components.}
\tablehead{\multicolumn{1}{c}{Field} &  \multicolumn{3}{c}{Primary cold component\tablenotemark{a}}  &  \multicolumn{3}{c}{Secondary cold component\tablenotemark{b}} & \multicolumn{1}{c}{Field Location}  \\
 & \multicolumn{1}{c}{$\langle v \rangle^{\rm primary}$} & \multicolumn{1}{c}{$\sigma_v ^{\rm primary}$}  &  \multicolumn{1}{c}{$N_{\rm primary}/N_{\rm total}$} &  \multicolumn{1}{c}{$\langle v \rangle^{\rm secondary}$} &   \multicolumn{1}{c}{$\sigma_v ^{\rm secondary}$}  &  \multicolumn{1}{c}{$N_{\rm secondary}/N_{\rm total}$} &  }
\startdata
f207 & $-524.4^{+7.7}_{-7.4}$ & $23.2^{+7.2}_{-5.0}$ &  $0.31^{+0.08}_{-0.07}$ & $-425.6^{+10.0}_{-8.3}$ &  $20.8^{+12.9}_{-7.6}$ & $0.31^{+0.11}_{-0.10}$ & GSS core \\
H13s & $-489.7^{+4.5}_{-4.2}$ & $21.3^{+4.0}_{-3.2}$ & $0.48^{+0.07}_{-0.06}$ & $-388.3^{+5.4}_{-5.4}$ & $17.0^{+6.6}_{-6.9}$ & $0.27^{+0.07}_{-0.08}$ & GSS core \\ 
a3 & $-440.5^{+5.2}_{-4.9}$ & $16.8^{+4.6}_{-3.3}$ & $0.59^{+0.11}_{-0.12}$ & ... & ... & ... & GSS core \\
d1 & $-382.0^{+12.4}_{-11.6}$ & $30.3^{+13.3}_{-8.6}$ & $0.57^{+0.15}_{-0.17}$ & $-289.2^{+6.8}_{-7.4}$ & $8.2^{+8.6}_{-4.5}$ & $0.15^{+0.11}_{-0.09}$ & west of highest-surface  \\
 & & & & & & & brightness region of GSS core \\
a13 & $-300.7^{+12.0}_{-11.6}$ & $32.2^{+11.8}_{-10.6}$ & $0.72^{+0.15}_{-0.21}$ & ... & ... & ... & GSS envelope  \\
m4 & $-355.3^{+5.1}_{-4.5}$ & $11.4^{+5.2}_{-4.1}$ & $0.45^{+0.14}_{-0.13}$ &  $-254.8^{+5.9}_{-4.9}$  & $6.6^{+6.0}_{-2.9}$ & $0.16^{+0.10}_{-0.08}$ & 60~kpc minor-axis \\
& & & & & & & arc (stream C)  \\
\enddata
\tablenotetext{a}{Quoted errors are the 90\% confidence limits from the maximum likelihood analysis.  All fields are assumed to also have a broad underlying distribution of stars with parameters $\langle v \rangle^{\rm sph} = -300$~\kms\ and $\sigma_v ^{\rm sph}=129$~\kms\ \citep{gilbert2007}.}
\tablenotetext{b}{In addition to the primary kinematically cold component, fields f207, H13s, d1 and m4 contain second kinematically cold components.}
\label{table:gss_mlfits}
\end{deluxetable*}
\def\arraystretch{1.0}


\subsection{GSS Core Fields}\label{sec:ongss}

\subsubsection{New GSS Field at $R\sim 17$~kpc}\label{sec:ongss_f207}

\begin{figure}[tb]
\plotone{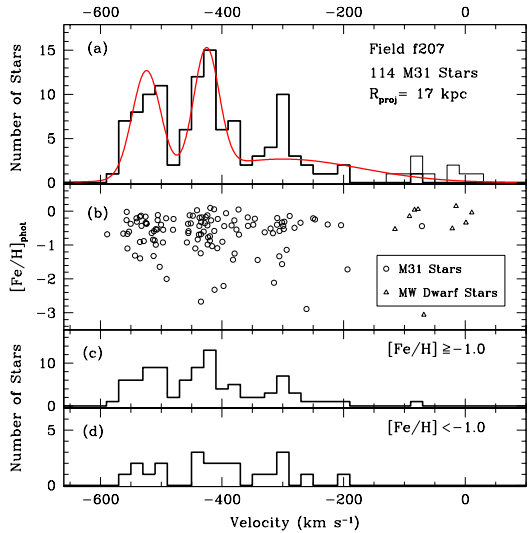}
\caption{Line-of-sight, heliocentric velocities of stars in the GSS core field at \rproj\,$=17$~kpc (field f207; \S\,\ref{sec:ongss_f207}).
({\it a}) The bold histogram denotes the distribution of stars classified as M31 red giants ( \S\,\ref{sec:gss_method}).  The red curve is the maximum-likelihood multi-Gaussian fit to the M31 RGB stars  (Table~\ref{table:gss_mlfits}).  The thin histogram denotes the velocity distribution of all stars (M31 RGB and MW dwarfs).
({\it b}) Velocity versus metallicity (\fehp) of the stars in field f207.  Stars classified as M31 RGB stars are denoted by circles and stars classified as MW dwarf stars are denoted by triangles.  
({\it c}) Velocity distribution of M31 RGB stars with \fehp\,$\ge -1.0$.  
({\it d}) Velocity distribution of M31 RGB stars with \fehp\,$<-1.0$.  
The GSS core is evident as a kinematically-cold peak of relatively metal-rich stars at $v\sim -525$~\kms\ in panels ({\it a})\,--\,({\it c}), and a second kinematically-cold, relatively metal-rich group of stars is seen at $v\sim -425$~\kms.  In addition, stars associated with M31's spheroid are present with a wide range of velocities centered on the systemic velocity of M31 ($v_{\rm sys}=-300$~\kms).
}
\label{fig:f207}
\end{figure}


Field f207 is located on the high-surface brightness core of the GSS at a 
projected radius of 17~kpc from the center of M31 (Fig.~\ref{fig:gss_roadmap}).   
The line-of-sight velocity distribution of stars in this field is shown in 
Figure~\ref{fig:f207}({\it a}), while panels ({\it b\,--\,d}) show the velocity 
distribution of stars in field f207 as a function of \fehp\ (\S\,\ref{sec:gss_data_met}).  
The GSS is the kinematically-cold, relatively metal-rich group of stars at very 
negative velocities ($v<-480$~\kms).  This is consistent with the expected velocity 
of the GSS at this position based on previous kinematical studies 
\citep[Fig.~\ref{fig:gss_velhist};][]{ibata2004,kalirai2006gss,guhathakurta2006}.  
The data in this field provide the innermost secure kinematical detection of the GSS 
to date.  

In addition to the GSS, a second prominent kinematically cold, metal-rich 
population can be seen in the 
velocity range $-460\lesssim v \lesssim -400$~\kms.  A less prominent, 
kinematically cold, primarily metal-rich group of stars may be 
present at $v\sim -300$~\kms.   As expected for a field
at this projected radial distance from M31, the majority of the stars not in
the kinematically cold peaks are also relatively metal-rich 
\citep[e.g.,][]{durrell2001,durrell2004,kalirai2006halo,gilbert2007}.  

The data are consistent (via a reduced-$\chi^2$ test) with being drawn from the 
maximum-likelihood Gaussian with three kinematical components: the GSS 
core at $v=-524$~\kms, a second kinematically cold component at 
$v=-426$~\kms, and M31's kinematically broad spheroid 
[Table~\ref{table:gss_mlfits}; Fig.~\ref{fig:f207}({\it a}), solid curve]. 
We therefore adopt the triple-Gaussian fit for this field in the analysis 
presented here.  Future observations may increase the statistical 
significance of the $v\sim -300$~\kms\ component, which currently has 
an excess of only $\sim 6$ stars ($\sim 5$\% of the M31 RGB stars in this field)
 above the number expected from 
M31's broad spheroid.  The stars in this 
possible kinematical peak have a mean velocity of $v\sim-305$~\kms\ and a 
dispersion of $\sim 8$~\kms. 

The measured velocity dispersion of the GSS core and the 
second kinematically cold component are 
$\sigma_v^{\rm GSS}=23.2^{+7.2}_{-5.0}$~\kms\ and 
$\sigma_v^{\rm sec}=20.8^{+12.9}_{-7.6}$~\kms, respectively. 
The median velocity error of M31 RGB stars in this field is 
6.9~\kms\ (\S\,\ref{sec:obs}). 
Thus the intrinsic velocity dispersions of the GSS and second 
kinematically cold component are $\sigma_v^{intr}=22.2$ and 
$\sigma_v^{intr}=19.6$~\kms, respectively.  We will present evidence in
(\S\,\ref{sec:prop_sec}) that the second kinematically
cold component may itself be a part of the GSS.

\subsubsection{Existing GSS Core Fields at 21~kpc and 33~kpc}\label{sec:ongss_H13s_a3}

\begin{figure}[tb]
\plotone{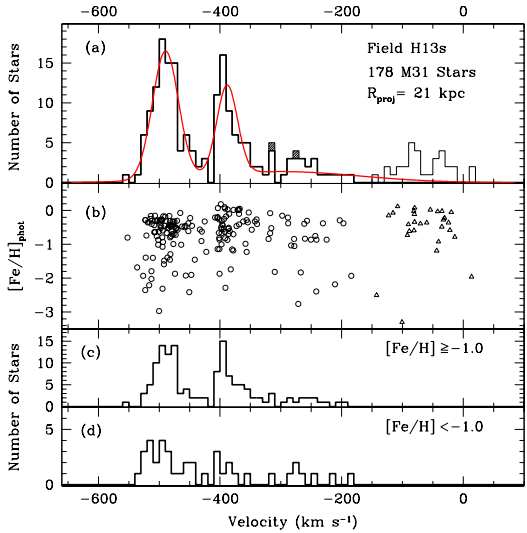}
\caption{
Line-of-sight, heliocentric velocities of stars in the GSS core field at \rproj\,$=21$~kpc (field H13s; \S\,\ref{sec:ongss_H13s_a3}).  Panels and symbols are the same as in Figure~\ref{fig:f207}.  The primary (GSS core) and the secondary kinematically cold components at $v\sim -490$~\kms\ and $v\sim -390$~\kms, respectively, are clearly seen in panels ({\it a})\,--\,({\it c}), as well as stars related to M31's spheroid with a wide range of velocities.  The shaded boxes in panel ({\it a}) denote stars that have physical characteristics that are consistent with the M31 RGB star population, but whose very blue \vio\ colors cause them to be classified as MW dwarf stars (\S\,\ref{sec:gss_method}).
}
\label{fig:H13s}
\end{figure}

\begin{figure}[tb]
\plotone{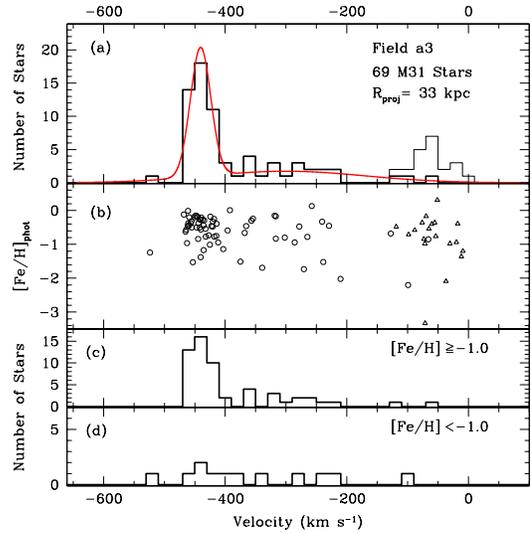}
\caption{
Line-of-sight, heliocentric velocities of stars in the GSS core field at \rproj\,$=33$~kpc (field a3; \S\,\ref{sec:ongss_H13s_a3}).  Panels and symbols are the same as in Figure~\ref{fig:f207}.  The GSS core is evident as a relatively metal-rich, kinematically cold population of RGB stars at $v\sim-440$~\kms\ in panels ({\it a})\,--\,({\it c}).  The velocity distribution of field a3 also has a significant contribution from M31 spheroid stars spanning a range of velocities and metallicities.
}
\label{fig:a3}
\end{figure}

Fields H13s and a3 are located near the eastern-edge of the GSS core at 
21 and 33~kpc (Fig.~\ref{fig:gss_roadmap}). 
Figures~\ref{fig:H13s} and \ref{fig:a3} display the line-of-sight velocity 
distributions of stars in these fields.  
Data in H13s and a3 have previously been presented by 
\citet{kalirai2006gss} and \citet{guhathakurta2006}, respectively.  
The data presented in this work benefit from
several improvements in our reduction and analysis procedures,
including an improved stellar template library for determining line-of-sight 
velocities, a correction to the measured velocity for mis-centering of the 
star in the slit, and the recovery of data from several CCD chips on the 
H13s\_2 mask which had initially failed our automated reduction pipeline 
(\S\,\S\,\ref{sec:obs}--\,\ref{sec:prev_pub}).  Our revised estimates presented
below for the mean velocity of the kinematically cold components in fields 
H13s and a3 are consistent with the $\sim +20$~\kms\ shift we have observed 
between our improved and original reduction pipelines 
(Fig.~\ref{fig:gss_velhist}).

The two kinematically cold, metal-rich components previously identified 
in field H13s are clearly seen in Figure~\ref{fig:H13s}, along with a kinematically
broad, relatively metal-rich spheroid component. The parameters for the 
maximum-likelihood triple Gaussian fit to the velocity distribution in 
this field are listed in Table~\ref{table:gss_mlfits}.  
 
Although both the GSS core at $v\sim -490$~\kms\ and the second kinematically 
cold component ($v\sim -390$~\kms) in 
field H13s are primarily metal-rich [Fig.~\ref{fig:H13s}({\it b,c})], they are  
also present in the velocity distribution of stars with 
\fehp\,$<-1.0$ [Fig.~\ref{fig:H13s}({\it b,d})].  
In fact, the GSS and second kinematically cold component comprise an 
estimated $51$\% and 19\% of the M31 RGB population with \fehp\,$<-1.0$, 
respectively, similar to the respective fractions estimated using the full velocity
distribution in H13s (Table~\ref{table:gss_mlfits}).   These percentages were 
estimated using a maximum-likelihood fit to the \fehp\,$<1.0$ velocity 
distribution in Fig.~\ref{fig:H13s}({d}), with all parameters except 
the fraction of stars in each component held fixed at the values in 
Table~\ref{table:gss_mlfits}.  

\citet{kalirai2006gss} measured dispersions of 
$\sigma_v=19$~\kms\ for both kinematically cold components in this field, which is in agreement with our new measurements (Table~\ref{table:gss_mlfits}).
The median velocity measurement error is 5.4~\kms\ in field H13s, 
yielding intrinsic velocity dispersions of $\sigma_v^{intr}=20.6$~\kms\ 
and $\sigma_v^{intr}=16.1$~\kms\ for the GSS core and second 
kinematically cold component, respectively.  

The prominent kinematically cold, metal-rich peak in the velocity 
distribution of field a3, reported as a kinematical detection of the 
GSS by \citet{guhathakurta2006}, is evident in Figure~\ref{fig:a3} 
along with a significant kinematically broad spheroid component that is
more metal-poor, as expected for M31 RGB stars at
this distance from M31 \citep{kalirai2006halo}.   Table~\ref{table:gss_mlfits}
lists the parameters from the maximum-likelihood double Gaussian fit to
the velocity distribution. 

\citet{guhathakurta2006} measured $\sigma_v^{\rm GSS}=21\pm 7$~\kms\ for the GSS in this field, consistent with our new measurement of $\sigma_v^{\rm GSS}=16.8^{+4.6}_{-3.3}$~\kms.  The median velocity measurement error in this field is 6.8~\kms, yielding an intrinsic velocity dispersion of $\sigma_v^{intr}=15.4$~\kms\ for the GSS core. 
 
\subsection{Fields Off the GSS Core}\label{sec:offgss}

\subsubsection{Fields to the West of the GSS Core}\label{sec:offgss_west}

\paragraph{South Quadrant Dwarf Spheroidal Field at R$\sim 45$~kpc}
Field d1 is located to the west of the highest surface-brightness region 
of the GSS (Fig.~\ref{fig:gss_roadmap}), yet is still solidly within the 
region of M31's spheroid dominated by the GSS (as evidenced by the
deeper photometric study of \citet{ibata2007}).  Although it was designed 
to study the dwarf spheroidal galaxy And~I, we have identified a total of 
48 M31 field stars in addition to the large number of dSph members in 
this field (\S\,\ref{sec:and_fields}).  

\begin{figure}[tb]
\plotone{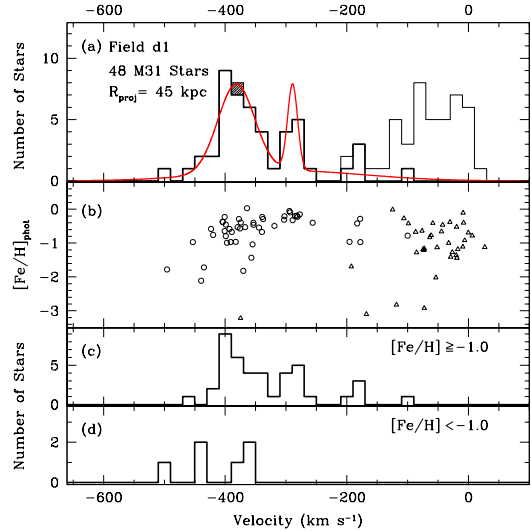}
\caption{
Line-of-sight, heliocentric velocities of stars in field d1, located in the southern quadrant of M31's halo at \rproj\,$=45$~kpc (\S\,\ref{sec:offgss_west}).  Panels and symbols are the same as in Figure~\ref{fig:f207} and the shaded boxes in panel~({\it a}) denote very blue stars as in Figure~\ref{fig:H13s}.  
Only stars that are judged unlikely to be associated with the dSph galaxy And~I are shown here (\S\,\ref{sec:and_fields}).  
The maximum-likelihood triple-Gaussian fit (Table~\ref{table:gss_mlfits}) is shown in panel ({\it a}). 
}
\label{fig:vf_and1}
\end{figure}

The line-of-sight velocity distribution of M31 field stars in field d1 
is shown in Figure~\ref{fig:vf_and1}.  There is a kinematical peak of 
relatively metal-rich stars at $v\sim -375$~\kms, and a 
compact group of $\sim 9$ stars ($\sim 19$\% of the M31 RGB stars in this 
field) in the $v$\,--\,\fehp\ plane 
at $v\sim -290$~\kms\ that are even more metal-rich 
[Fig.~\ref{fig:vf_and1}({\it b})].  The spread in metallicity of the stars 
at $v\sim-290$~\kms\ is only $0.1$~dex.  

M31's halo has been 
shown to become increasingly metal-poor towards its outskirts.  
The median \fehp\ of all the M31 RGB 
stars in this field is $-0.51$~dex, significantly higher than the expected 
metallicity of M31's halo 
at this projected distance 
\citep[\feh\,$\lesssim -1.0$;][]{kalirai2006halo,chapman2006,koch2008}.
The selection criteria to separate M31 field stars from dSph members 
only precludes stars with \fehp\,$<-1.0$ and $-425 < v < -375$~\kms\ from 
being included in the M31 RGB sample (\S\,\ref{sec:and_fields}), and thus is
not a significant factor in increasing the mean metallicity of stars in 
the field.   

The prominent kinematical peaks at $v\sim -380$~\kms and 
$v\sim -290$~\kms\ combined with the relatively high metallicity of the 
stars is strong evidence that we have detected substructure in this field.  
Furthermore, the data are strongly inconsistent with being drawn from a 
single broad ($\sigma_v\gtrsim100$~\kms) Gaussian centered on M31's systemic
velocity (with a probability $P_{\chi^2}\ll0.1$).  
The parameters of the maximum-likelihood triple-Gaussian fit to the M31 RGB 
stars in this field are listed in Table~\ref{table:gss_mlfits}.  A 
maximum-likelihood fit with the dispersion of M31's spheroid set to the value 
expected for \rproj$\sim 45$~kpc by the \citet{chapman2006} study 
(\S\,\ref{sec:cks}) returns parameters that are 
within the 90\% confidence limits of the fit in Table~\ref{table:gss_mlfits}.   

The location of field d1 (Figs.~\ref{fig:gss_roadmap} \& 
\ref{fig:gss_velhist}) argues strongly for the fact that the GSS is the source
of at least some of the kinematic substructure seen in this field.  The 
$v\sim -380$~\kms\ kinematically cold
component has the appropriate mean velocity to be the GSS 
(Fig.~\ref{fig:gss_velhist}), and we will show 
in \S\,\ref{sec:prop_core} that the kinematical and chemical properties of 
the $v\sim-380$~\kms\ feature are consistent with this interpretation. 
The nine stars with velocities $\sim-290$~\kms\ and \fehp\,$\sim-0.2$~dex 
are evenly spread over the regions of spectroscopic masks in which we find 
M31 RGB field stars.  They thus represent a kinematically cold tidal
stream, not a spatially compact group of stars such as an intact cluster.  

\paragraph{South Quadrant M31 Halo Field at $R\sim 58$~kpc}
Field a13 is located far to the west of the high-surface brightness 
core of the GSS at \rproj\,$=58$~kpc, solidly within the the prominent ``GSS 
envelope'' photometric structure (Fig.~\ref{fig:gss_roadmap}).  
The surface brightness of this field, estimated from star counts of M31 
RGB and Milky Way dwarf stars, is significantly ($>3\sigma$) brighter 
than expected based on the surface brightness profile of M31's stellar 
halo derived from star counts in fields from 18~kpc to 165~kpc 
\citep{guhathakurta2005}.  This implies that field a13 likely contains 
a significant amount of substructure which boosts its surface brightness 
above what is expected for a relatively smooth halo field at a radial 
distance of 58~kpc.  

\begin{figure}[tb]
\plotone{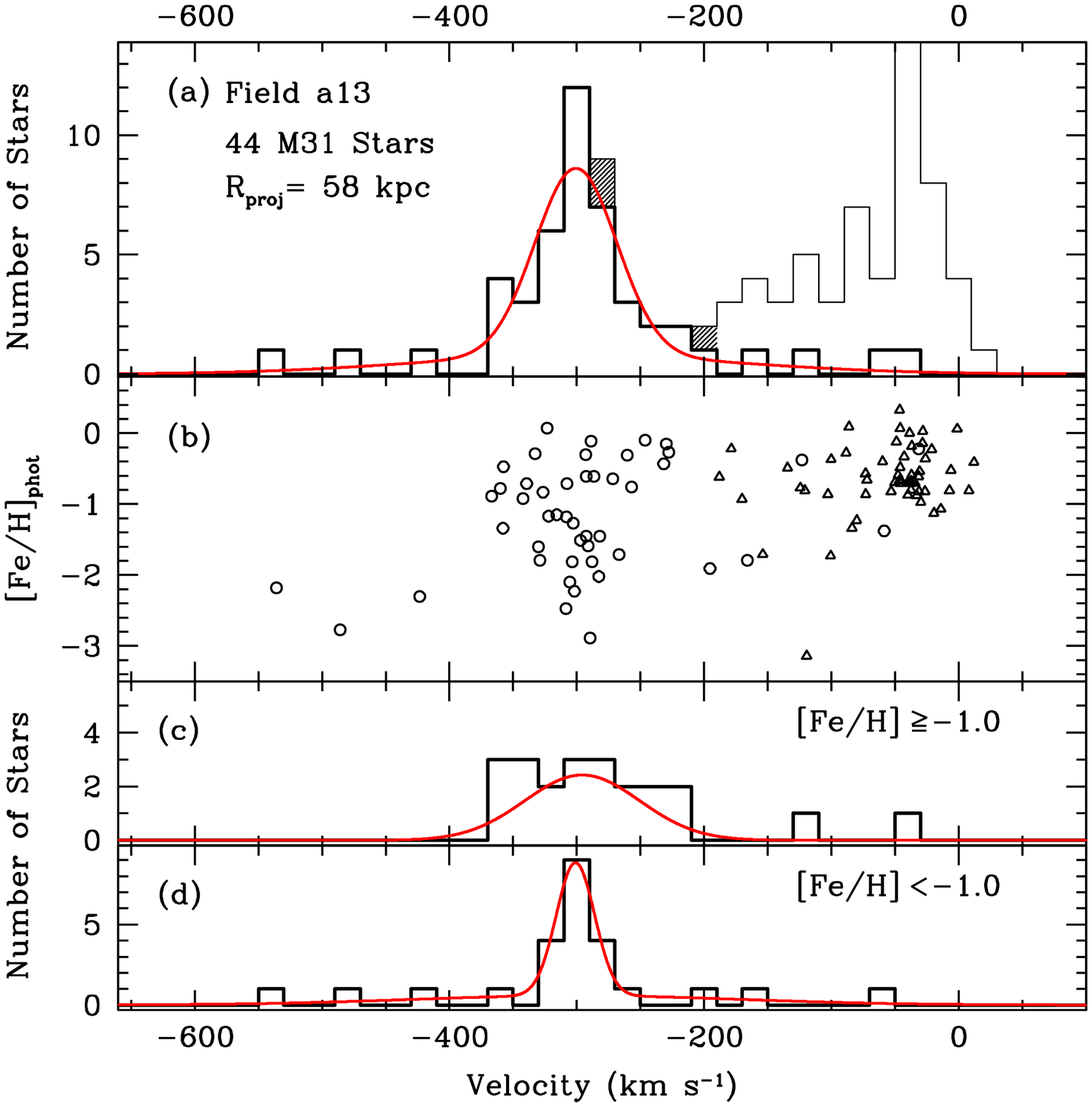}
\caption{
Line-of-sight, heliocentric velocities of stars in the south quadrant M31 halo field at \rproj\,$=58$~kpc (field a13; \S\,\ref{sec:offgss_west}) with the maximum-likelihood double-Gaussian fit (red curve) overlaid.  Panels and symbols are the same as in Figure~\ref{fig:f207} and the shaded boxes in panel~({\it a}) denote very blue stars as in Figure~\ref{fig:H13s}.  The velocity and \fehp\ distribution of M31 RGB stars in this field shows evidence of multiple kinematical components.  Relatively metal-poor M31 spheroid stars with a wide range of velocities are present, as well as a colder kinematical component centered at $v\sim -300$~\kms\ [panels~({\it b})\,--\,({\it d})]. The relatively metal-rich subset [\fehp\,$\ge -1.0$, panel ({\it c})] of the stars with $v\sim -300$~\kms\ have a somewhat broader velocity dispersion than the relatively metal-poor subset [\fehp\,$< -1.0$, panel ({\it d})]. 
}
\label{fig:a13}
\end{figure}

The line-of-sight velocity distribution of stars in this field is 
displayed in Figure~\ref{fig:a13}.  Stars are present with a wide range of 
metallicity.  The relatively metal-rich stars (\fehp\,$\ge -1.0$) span a 
fairly limited range of velocities ($ -370 \lesssim v \lesssim - 230$~\kms), 
while the relatively metal-poor stars (\fehp\,$<-1.0$) display both a 
wide range of line-of-sight velocities ($-550\lesssim v \lesssim -150$~\kms) 
and a peak of stars with a narrower velocity dispersion centered at 
$v\approx -300$\kms.    


The parameters from the maximum-likelihood double-Gaussian fit to the 
full velocity distribution are listed in Table~\ref{table:gss_mlfits}.  
If a smaller velocity dispersion of $\sigma_v=100$~\kms\ is assumed for M31's 
spheroid, the resulting maximum-likelihood parameters lie 
well within the 90\% confidence limits 
shown in Table~\ref{table:gss_mlfits}.  The kinematically cold component 
is estimated to comprise a large fraction ($\sim 70$\%) of the stars in 
this field.  If the number of stars in M31's smooth spheroid
is assumed to be 30\% of the M31 RGB stars in this field, the 
estimated spheroid surface brightness for field a13 
is fully consistent with the  observed surface brightness profile of M31's 
stellar halo.

The kinematical data are inconsistent ($P_{\chi^2}\ll0.1$\%) with being 
drawn from a relatively broad ($\sigma_v\geq 100$~\kms) single Gaussian 
centered on M31's velocity, but are marginally consistent with the 
maximum-likelihood single-Gaussian fit ($\sigma_v=74^{+15}_{-11}$~\kms, 
$P_{\chi^2}\sim 6$\%).  However the distribution 
of the data in the \fehp\,--\,$v$ plane, the enhanced surface brightness 
observed in this field, and its location on a known photometric substructure 
(the GSS envelope) provide convincing evidence that there 
is another RGB population in this field in addition to M31's kinematically 
broad spheroid.  The field's location in the photometrically 
prominent GSS envelope makes this the most likely source of the 
substructure seen in field a13.  

The maximum-likelihood double-Gaussian fit to all the stars in this
field is a simplification of the distribution of the data.
Since one of the Gaussian components is held fixed with parameters 
representing M31's kinematically broad, smooth spheroid population, only 
one Gaussian component is available to fit the triangular 
distribution seen in Figure~\ref{fig:a13}({\it b}).  Panels ({\it c}) 
and ({\it d}) show quite clearly the difference in the velocity distributions 
of stars with \fehp$\geq-1.0$ and \fehp$<-1.0$~dex.  Gaussian fits to the 
velocity distribution of the stars in each of these panels have been overlaid.
We assume a broad kinematical component is present in the \fehp$<-1.0$ 
population, but not in the \fehp$\geq-1.0$ population.  This is motivated in 
part by the velocity distribution of the stars [there are clearly metal-poor 
stars spread over a very large range in velocity in Fig.~\ref{fig:a13}({\it d}) while a similar distribution is not seen in Fig.~\ref{fig:a13}({\it c})] 
and in part by the fact that M31's stellar halo has been shown to be 
largely metal-poor at these large radii 
\citep[$\langle$\,\fehp$\rangle\sim-1.5$][]{kalirai2006halo,koch2008}.  
The velocity dispersion of the kinematical peak of stars with \fehp$\geq-1.0$
is $\sigma_v=45^{+15}_{-10}$~\kms, and the velocity dispersion of the kinematical peak 
of stars with \fehp$<-1.0$ is $\sigma_v=15.1^{+7.4}_{-5.2}$~\kms.  These values are 
both marginally consistent with the dispersion found in the 
fit to the full velocity distribution (Table~\ref{table:gss_mlfits}).  
The mean velocities are fully consistent with each other 
and with the results of the fit to the full velocity distribution.

We can only speculate for now on the relationship between the relatively 
metal-poor, narrower kinematical peak seen in Figure~\ref{fig:a13}({\it d}) 
and the relatively metal-rich, slightly broader kinematical peak in 
Figure~\ref{fig:a13}({\it c}).  Another pointing in a different region 
of the GSS envelope is necessary to determine if this structure in 
[Fe/H]\,--\,$v_{\rm los}$ space persists throughout the GSS envelope region, 
or is specific to the location of field a13.  If it persists, the interesting 
structure seen in the \fehp\,--\,$v$ is likely endemic to the stars associated 
with the GSS envelope and represents one coherent substructure.  
If only part of this structure is seen in another location on the GSS envelope, 
(i.e., only the relatively metal-poor, narrower kinematical peak or only 
the relatively metal-rich, slightly broader kinematical peak) it would 
give credence to the interpretation of this structure as two 
independent kinematical substructures in M31's halo which are superimposed 
along our line-of-sight.  For the present, we use the fit to the full sample
of stars in this field (Table~\ref{table:gss_mlfits}), and treat the 
substructure as one (complex) kinematical component, identified as the first 
kinematical detection of the GSS envelope.

\paragraph{South Quadrant Dwarf Spheroidal Field at R$\sim 69$~kpc}
Field d3 is far to the west of the high-surface brightness GSS core, 
and on the edge of the region of M31's stellar halo which appears 
contaminated by the GSS in starcount maps (Fig.~\ref{fig:gss_roadmap}). 
Although the spectroscopic masks were designed to study RGB members of the 
dwarf spheroidal galaxy And~III, we have also identified a total of 13 M31 
field stars (\S\,\ref{sec:and_fields}).   

\begin{figure}[tb]
\plotone{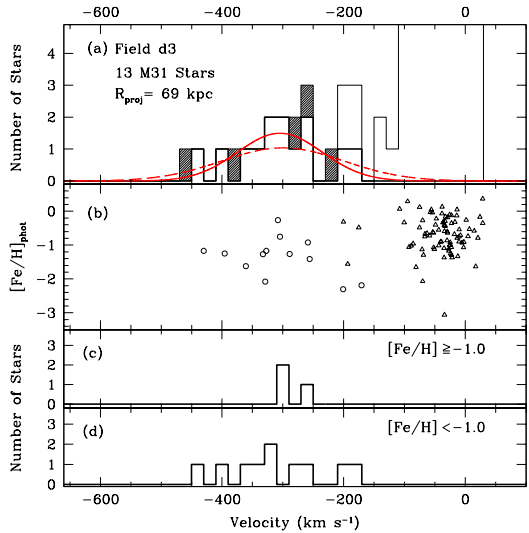}
\caption{
Line-of-sight, heliocentric velocities of stars in the dwarf spheroidal field d3, located in the southern quadrant of M31's halo at \rproj\,$=69$~kpc (\S\,\ref{sec:offgss_west}).  Panels and symbols are the same as in Figure~\ref{fig:f207} and the shaded boxes in panel~({\it a}) denote very blue stars as in Figure~\ref{fig:H13s}.  
Only stars that are unlikely to be associated with the dSph galaxy And~III are shown here (\S\,\ref{sec:and_fields}).  
The maximum-likelihood single Gaussian fit to the data in this field has parameters $\langle v \rangle=-305$, $\sigma_v=69.4$~\kms\ (solid curve).  A Gaussian with parameters $\langle v \rangle=-300$, $\sigma_v=100$~\kms\ is also shown (dashed curve). 
}
\label{fig:vf_and3}
\end{figure}

The line-of-sight velocity distribution of the 13 M31 field stars in field 
d3 is shown in Figure~\ref{fig:vf_and3}.  The M31 RGB stars span a 
wide range in velocity and \fehp\ values, and the velocities are consistent 
with being drawn from either a Gaussian with parameters \mvsph\,$=-300$~\kms, 
\sigvsph\,$=129$~\kms\ or \sigvsph\,$=100$~\kms\ [the dispersion implied 
for the projected radius of this field by the \citet{chapman2006} study].  
The maximum-likelihood single Gaussian in this field has parameters 
$\langle v \rangle=-304.8^{+33.3}_{-33.8}$, 
$\sigma_v=69.4^{+30.2}_{-17.5}$~\kms.  Based on the current limited sample of 
stars, there is no kinematical evidence of substructure in this field. 

\subsubsection{South-East Minor Axis Field at $R\sim 57$~kpc}\label{sec:m4}

\begin{figure}
\plotone{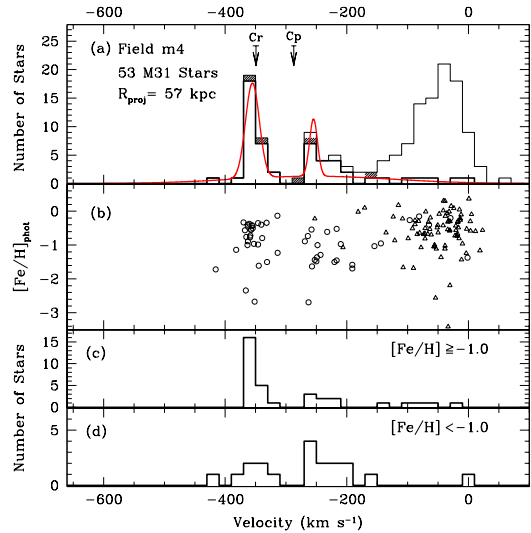}
\caption{
Line-of-sight, heliocentric velocities of stars in the southeast minor axis field at \rproj\,$=57$~kpc (field m4; \S\,\ref{sec:m4}).  Panels and symbols are the same as in Figure~\ref{fig:f207}.  There is a prominent, kinematically cold peak of metal-rich stars in this field centered at $v\sim-360$~\kms, and a second kinematically cold peak of more metal-poor stars at $v\sim-260$~\kms.  The arrows denote the mean velocities and dispersions of kinematically cold features detected in a nearby spectroscopic field \citep{chapman2008}. 
As in Figure~\ref{fig:H13s}, shaded boxes in panel~({\it a}) denote blue stars for which our M31 RGB/MW dwarf classification is less certain (\S\,\ref{sec:gss_method}).  
}
\label{fig:m4}
\end{figure}

Field m4 is located on M31's southeast minor axis at \rproj$\sim 57$~kpc (Fig.~\ref{fig:gss_roadmap}), at the 
same location as the arc-like photometric substructure termed `stream C'.  The line-of-sight 
velocity distribution of stars in this field is displayed in Figure~\ref{fig:m4}.  
Two prominent kinematically cold features are present: a 
relatively metal-rich (\fehp\,$\gtrsim -1.0$) group of stars at $v\sim -350$~\kms\ 
and a secondary group of stars with somewhat lower metallicities at $v\sim -260$~\kms.
Another spectroscopic study of M31 RGB stars $\sim 1$\degree\ southwest of 
field m4 also found evidence of two kinematically-cold components 
\citep[][the mean velocities and dispersions of these components are designated 
by the arrows in Fig.~\ref{fig:m4}]{chapman2008}.  In \S\,\ref{sec:prop_maxis} we 
will show that both kinematically cold components 
can be identified with photometric substructure seen in the region of stream C. 

The parameters of the maximum-likelihood triple-Gaussian fit to stars in field m4 are 
presented in Table~\ref{table:gss_mlfits}.  The maximum-likelihood triple Gaussian 
parameters assuming a smaller velocity dispersion for M31's spheroid ($\sigma_v=100$~\kms) 
are well within the 90\% confidence limits in Table~\ref{table:gss_mlfits}.  
The median velocity measurement error in field m4 is 5.4~\kms.  Thus the intrinsic velocity 
dispersions of the primary and secondary kinematically cold components are 
$\sigma_v^{\rm intr}=10.0$~\kms\ and 3.8~\kms, respectively; the secondary 
kinematically-cold component is just resolved.  

\subsection{Summary of the Results for Individual Fields}

We have detected and measured the mean velocity and velocity dispersion 
of the GSS in 5 fields: f207, H13s, a3, d1, and a13.  Of these, field 
f207 provides the innermost 
measurement of the mean velocity and velocity dispersion of the GSS at 
\rproj\,$\sim 17$~kpc, and field a13 provides the first measurement of the 
kinematical properties of the GSS envelope.   Four of our 7 fields show not 
one but {\it two} kinematically cold components
in addition to the kinematically hot M31 spheroid population. 
Field f207 has a second kinematically cold component with a mean velocity 
$\sim 100$~\kms\ more positive than that of the GSS
(\S\,\ref{sec:ongss_f207}).  
This component mirrors the secondary
kinematically cold component in field H13s, which also has a mean velocity 
$\sim 100$~\kms\ more positive than the GSS's.  The second kinematically-cold 
components in fields f207 and H13s likely
have the same physical origin as each other, and may also be part of the GSS 
(\S\,\ref{sec:prop_sec}). 
The velocity distribution of field m4 shows two prominent, kinematically-cold 
peaks (\S\,\ref{sec:m4}).  They confirm a previous finding of two 
kinematically cold components in the 
photometrically-identified minor-axis arc stream C.   

\section{A Comparison of the Kinematical and Chemical Abundance Properties of the GSS and Other Debris}\label{sec:prop}
In \S\,\ref{sec:ind_fields}, we identified and characterized kinematically cold components
in individual fields.  In this section, we widen our focus and discuss the chemical 
and kinematical properties of these components in relation to each other and in 
the context of physical structures known to exist in M31's spheroid.  
We also examine the evidence for various physical origins for the kinematical components 
which are not clearly attributable the the GSS: the kinematically cold components in field m4 (\S\,\ref{sec:prop_maxis}) and the second kinematically cold components in fields f207 and H13s (\S\,\ref{sec:prop_sec_data}).

\subsection{GSS Core and Envelope}\label{sec:prop_core}
Debris from the GSS extends over a large portion of the southern-most 
quadrant of M31's stellar halo.  The structure of the GSS includes the 
high-surface brightness, high metallicity 
``core''  
($\langle$\,[Fe/H]\,$\rangle = -0.54$) surrounded by a more diffuse, 
lower metallicity ``envelope'' ($\langle$\,[Fe/H]\,$\rangle = -0.71$) \citep{ibata2007}.  
Numerical simulations of the collision of the GSS's progenitor with M31 show 
that the core/envelope structure is naturally produced if the progenitor 
is a disk galaxy with an internal metallicity gradient \citep{fardal2008}.  

We have presented kinematical detections of the GSS core in fields f207 (\S\,\ref{sec:ongss_f207}), H13s and a3 (\S\,\ref{sec:ongss_H13s_a3}).  We have also detected kinematically cold components 
in fields d1 and a13 which we attribute to the GSS core and 
GSS envelope (\S\,\ref{sec:offgss_west}), respectively.  We will show 
below that their respective velocity and chemical properties corroborate 
this interpretation.

\subsubsection{Kinematics}\label{sec:prop_gss_kin}
\begin{figure}[tb]
\plotone{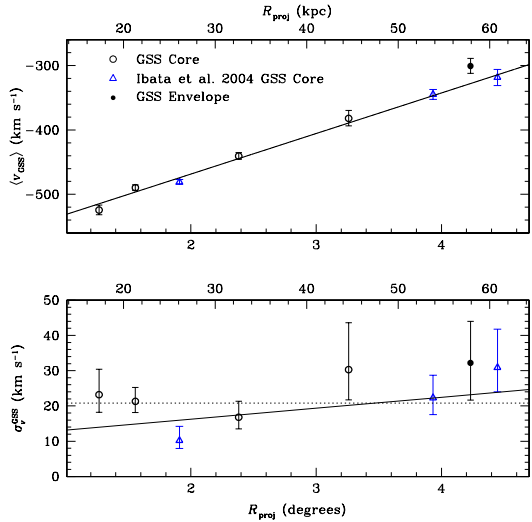}
\caption{Mean line-of-sight velocity and observed line-of-sight velocity 
dispersion (not corrected for measurement error) of stars associated with 
the GSS as a function of radial projected
distance from the center of M31 (\S\,\ref{sec:prop_gss_kin}).  The values are from maximum-likelihood, multiple-component Gaussian 
fits (\S\,\ref{sec:ind_fields}) to the stellar velocity distributions of 
each of the fields, including GSS core fields (open circles), the GSS 
envelope field (filled circle), and \citet{ibata2004} GSS core 
fields (open triangles).  Error bars denote 90\% confidence limits from 
the maximum-likelihood analysis.  Weighted least-squares linear fits to the 
GSS core data from fields f207, H13s, IF6, a3, IF2, and IF1 are also shown.  
The dotted line in the bottom panel shows the mean velocity dispersion of 
these fields.  The mean line-of-sight velocity and velocity dispersion of 
the GSS in fields d1 and a13 (open circle at 3.25\degree\ and solid circle, 
respectively) are consistent with the fits to the core fields along the 
eastern edge of the GSS. 
}
\label{fig:veltrends}
\end{figure}

An inspection of the velocity distributions of fields on the GSS 
clearly shows that its velocity 
(marked with arrows in Fig.~\ref{fig:gss_velhist}) 
becomes increasingly positive with increasing 
distance from the center of M31 
($\langle v\rangle_{\rm GSS}=-527$~\kms\ in field 
f207, while $\langle v\rangle_{\rm GSS}=-318$~\kms\ in field IF1).
Figure~\ref{fig:veltrends} displays the mean line-of-sight velocity and 
observed velocity dispersion (Table~\ref{table:gss_mlfits}) of the GSS as 
a function of projected radius from M31's center.   The mean velocity and 
velocity dispersion of the GSS core in the \citet{ibata2004} fields 
(Fig.~\ref{fig:gss_velhist}) were calculated using maximum-likelihood 
double Gaussian fits, with \Gsph\ held fixed with parameters 
$\langle v \rangle=-300$~\kms, $\sigma_v^{\rm sph}=129$~\kms\ 
(\S\,\ref{sec:cks}).  The error bars represent the 90\% confidence
limits from the maximum-likelihood analysis. 

We performed a weighted least-squares linear fit to the mean 
line-of-sight velocity of the GSS as a function of \rproj.  The result,
using the fields on the highest surface brightness region of the GSS core, is:
\begin{equation}
\langle v \rangle^{\rm GSS}= [ 63.0 (R_{\rm proj}/1^\circ)-594.4 ]~{\rm km s}^{-1}\\
\end{equation}
This fit includes the data from fields f207, H13s, IF6, a3, IF2, and IF1 
(Fig.~\ref{fig:gss_roadmap}).  These fields are represented by open 
circles and triangles in Figure~\ref{fig:veltrends}. The open circle at 
3.25\degree, field d1, was not included in the above fit and is discussed 
further below.  

Although the line-of-sight velocity of the orbit of the GSS progenitor is 
expected to be strongly non-linear over the range of radii observed, the GSS 
exists because the progenitor and stars belonging to the GSS are on slightly 
different orbits.  When this is taken into account in modelling the interaction 
which produced the GSS, the orbital models produce a roughly linear trend for 
the mean line-of-sight velocity of the GSS as a function of radial distance 
over most of the observed distance range \citep{fardal2006}. 
The linear approximation is an adequate description of the data over 
the range of radii for which the velocity of the GSS has been measured.

The d1 field (\rproj\,$\sim 45$~kpc) is very near the region chosen 
by \citet{ibata2007} for analysis of the GSS core's stellar population and 
is consistent with an extrapolation of this region towards M31's center 
(Fig.~\ref{fig:gss_roadmap}).  The GSS core velocity at the projected radius 
of our d1 field, based on the above linear fit, is expected to be 
$\langle v \rangle \sim -390$~\kms.  This is consistent with the velocity 
found for the primary kinematically cold component in field d1
(Table~\ref{table:gss_mlfits}, \S\,\ref{sec:offgss_west}).  

In addition to the fields in the GSS core, Figure~\ref{fig:veltrends} also 
displays the mean line-of-sight velocity and velocity dispersion of the 
kinematically cold component in the a13 field (filled circle).  The mean 
velocity of the kinematically cold component in field a13 is 
more positive than expected from the linear fit to the core fields.  The mean velocity of 
the GSS envelope is expected to be relatively close to, but not necessarily 
the same, as the mean velocity of the GSS core at the same projected radius 
(M.\ Fardal, private communication).  

There is not a significant trend with \rproj\ in the line-of-sight velocity 
dispersion of stars in the GSS (Fig.~\ref{fig:veltrends}, bottom panel), 
although the data hints that the velocity dispersion in the outer 
regions of the GSS (\rproj\,$>40$~kpc) may be slightly 
larger on average than the inner regions (\rproj\,$<40$~kpc).  The mean 
velocity dispersion of the GSS in the fields along its eastern edge is 
$\langle \sigma_v \rangle=20.8$~\kms.  
The velocity dispersion measured in the more western GSS core field d1 
is consistent with this mean within the 90\% confidence limits from the 
maximum-likelihood analysis, while the velocity dispersion of the GSS envelope 
field a13 is within the 99\% confidence limits.  The  
weighted least-squares fit to the eastern GSS core fields is also shown in
Figure~\ref{fig:veltrends}.  The velocity dispersion of the GSS in fields d1 and 
a13 is consistent with this linear fit. 

\subsubsection{Metallicity Distributions}\label{sec:prop_gss_met}

\begin{figure}[tb]
\plotone{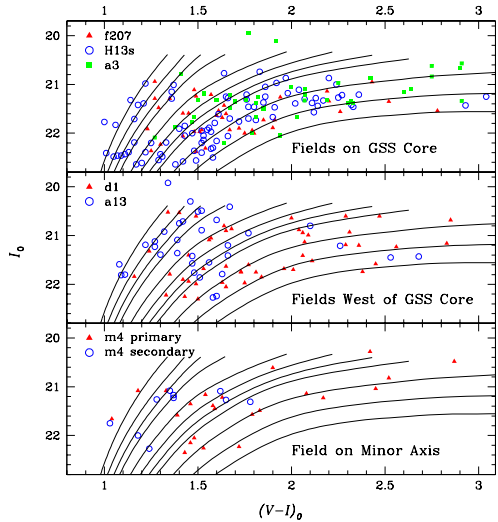}
\caption{Position of stars in the \ivio\ CMD for stars within $\pm 2\sigma_v$ of the mean velocity of the primary substructure in fields on and off the GSS core (\S\,\ref{sec:prop_gss_met}).  Isochrones shown are $12$~Gyr, [$\alpha$/Fe]$=0.0$ from \citet{vandenberg2006} and are the same set shown in Figure~\ref{fig:and1}.  The most metal-poor (bluest) isochrone represents [Fe/H]$=-2.31$ and the most metal-rich (reddest) isochrone represents [Fe/H]$=0.0$.  The CMD distribution of GSS stars in the three core GSS fields are similar, while the stars in the substructure in a13 and the secondary kinematically cold component in m4 are on average more blue.  
}
\label{fig:cmd_3pnl}
\end{figure}

Figure~\ref{fig:cmd_3pnl} shows the position in an \ivio\ CMD of all stars 
within $\pm 2\sigma_v$ of the mean velocity of the primary kinematical 
substructure in fields on and off the GSS core.  The 
stars within $\pm 2\sigma_v$ of the GSS core in fields f207, H13s, and a3 
have similar distributions in color-magnitude space.  The stars within $\pm 2\sigma_v$ 
of the kinematically cold component in field a13, located on the GSS envelope, 
are significantly bluer than the stars associated with the GSS core.
We have not corrected 
for the varying line-of-sight distance to the GSS in Figure~\ref{fig:cmd_3pnl}. 
The difference in the line-of-sight distance to the GSS between fields a3 and 
f207 is $\sim 20$~kpc \citep{mcconnachie2003}, which corresponds to only a 
0.05~dex difference in apparent magnitude.   

Differences in the CMD distributions of RGB stars are caused by differences 
in the metallicity, age, or alpha-enhancement of the stellar populations.  
For sufficiently old stellar populations 
\citep[older than $\sim 4$~Gyr;][]{gallart2005}, the RGB is primarily 
sensitive to metallicity, with age and alpha-enrichment playing secondary 
roles.  Although the alpha-enrichment of RGB stars in M31's spheroid and 
the GSS is not observationally constrained, these stars are known to span 
a range of ages (\S\,\ref{sec:gss_data_met}), in contrast to the 12~Gyr old 
coeval population we assume in measuring \fehp.  
With these caveats, we interpret differences in the CMD 
distribution of stars as differences in metallicity in the following 
analysis.  Due to the 
age-metallicity degeneracy of the RGB, the 
{\it differences} between the metallicity distributions of features, which 
reflect the observed differences in the CMD distributions, deserve more 
weight than the absolute metallicity estimates of the stars. 

The assumed distance to the stellar population also affects the photometric 
metallicity estimates of stars.  However, the line-of-sight distance to the
GSS core in individual fields is not known with a great deal of certainty, and 
the line-of-sight distances to the other kinematic features identified in
\S\,\ref{sec:ind_fields} are not yet known.  We therefore assume M31's distance
modulus when deriving the photometric metallicity estimates.
For the GSS kinematic features, we will note the effect 
of assuming the distance measured by \citet{mcconnachie2003} on the
\fehp\ value.       

The final source of systematic error in deriving the mean metallicity 
abundance of a kinematic feature is the assignment of individual stars 
to kinematic components.
Individual stars can be associated with the kinematically cold components 
only statistically (i.e., through maximum-likelihood Gaussian fits as 
in \S\,\ref{sec:ind_fields}).  Therefore, the [Fe/H] distributions discussed
below include stars associated with the kinematically cold components as 
well as stars associated with M31's spheroid that happen to have velocities 
within 
$\pm 2\sigma_v$ of the mean velocity of the kinematically cold components.  
For the GSS core fields, this is a relatively small effect since the GSS core 
is a narrow and dominant feature well offset from M31's systemic velocity 
(Fig.~\ref{fig:gss_velhist}).  

\begin{figure}[tb]
\plotone{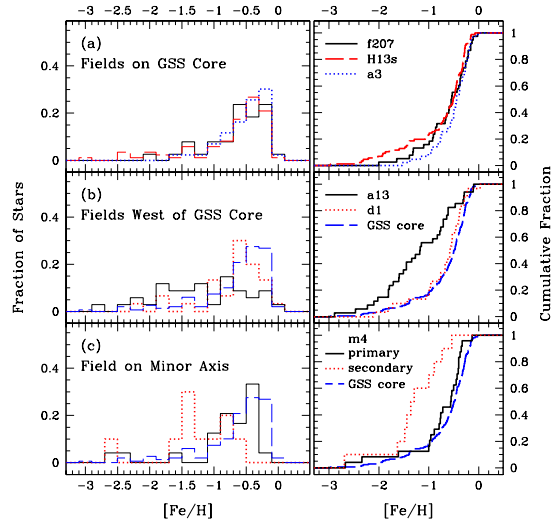}
\caption{Metallicity distribution function (MDF) of stars within  $\pm 2\sigma_v$ of the mean velocity of the primary substructure in fields on and off the GSS core, in histogram ({\it left}) and cumulative ({\it right}) form.   ({\it a}) MDF of the GSS core components in fields f207 (solid histogram and curve), H13s (dashed histogram and curve), and a3 (dotted histogram and curve), discussed in \S\,\ref{sec:prop_gss_met}.   ({\it b}) MDF of stars associated with the kinematically cold component in the d1 (dotted histogram and curve) and a13 (solid histogram and curve) fields.  The combined MDF of the GSS core components in panel ({\it a}) is shown for comparison (dashed histogram and curve, \S\,\ref{sec:prop_gss_met}).  ({\it c})  MDF of stars associated with the $v\sim-350$~\kms\ (solid histogram and curve) and $v\sim-250$~\kms\ (dotted histogram and curve) kinematically cold components in field m4 (\S\,\ref{sec:prop_maxis}).  As in panel ({\it b}), the MDF of the combined GSS core components is shown for comparison (dashed histogram and curve).  The GSS 
core MDFs from fields f207, H13s, and a3 are consistent with one other 
and are similar to those of the d1 field and
the m4 primary kinematically cold component, while the MDFs of the GSS 
envelope in field a13 and the secondary kinematically cold component in field m4 are 
significantly more metal poor.
}
\label{fig:mdf_gss}
\end{figure}

Figure~\ref{fig:mdf_gss} displays the [Fe/H] distributions of the stars in 
Figure~\ref{fig:cmd_3pnl} in differential and cumulative form.  The GSS core 
has a very similar [Fe/H] distribution in all three GSS core fields 
[Fig.~\ref{fig:mdf_gss}({\it a})], with a strong metal-rich peak and an 
extended metal-poor tail.  The median metallicity of stars within 
$\pm 2\sigma_v$ of the mean velocity of the GSS component is 
$\langle$[Fe/H]$\rangle_{\rm med}=-0.55$, $-0.51$, and $-0.44$ in fields 
f207, H13s, and a3, respectively; $\langle$[Fe/H]$\rangle_{\rm med}=-0.49$ 
($\langle$[Fe/H]$\rangle_{\rm mean}=-0.66\pm0.04$) for all three fields 
combined.  

Accounting for the estimated distance to the GSS in each of the above 
fields \citep{mcconnachie2003} results in only a $\sim -0.04$~dex difference 
in the derived \fehp\ estimates, while changing the assumption of the age of
the population from 12~Gyr to 8~Gyr results in a $\sim +0.1$~dex difference
in the derived \fehp\ estimates.   
Based on the multi-Gaussian fits, between 5\% and 8\% (or $\sim 3$ to 4 stars) 
of the red giants within $\pm 2\sigma_v$ 
of the mean velocity of the GSS core in fields f207, H13s, and a3 are 
associated with M31's spheroid.  Within \rproj\,$\sim 30$~kpc, M31's 
spheroid is found to be relatively metal-rich 
\citep{kalirai2006halo,gilbert2007,koch2008}.  However, if we assume 
that the GSS core is more metal-rich than M31's spheroid, and remove the 
3 to 4 most metal-poor stars in each field, 
the resulting median \fehp\ estimate for the GSS increases by only $0.02$~dex.

The [Fe/H] distribution of stars within $\pm2\sigma$ of the mean velocity 
of the $-380$~\kms\ kinematically cold component in field d1 is very 
similar to the [Fe/H] distribution of stars associated with the GSS core 
[Fig.~\ref{fig:mdf_gss}({\it b})].  The median metallicity of the stars associated with
the GSS in field d1 is $\sim 0.1$~dex lower than the median [Fe/H] of the 
GSS core in fields f207, H13s, and a3. 
If the estimated distance to the GSS in each of the fields is accounted for, 
the median metallicity of 
the GSS in field d1 is found to be $0.13$~dex more metal-poor than in the more eastern 
GSS core fields f207, H13s and a3.

Stars within $\pm 2\sigma_v$ of the mean velocity of the kinematically 
cold component in field a13 are on average significantly more metal-poor than 
the GSS core [Fig.~\ref{fig:mdf_gss}({\it b})]. 
The median metallicity of the stars associated with the 
kinematically cold component in field a13 is $\sim 0.7$~dex lower than 
the median [Fe/H] of the GSS core ($\langle$[Fe/H]$\rangle_{\rm med}=-1.18$, $\langle$[Fe/H]$\rangle_{\rm mean}=-1.19\pm0.12$).   
This is qualitatively consistent with results from purely photometric data that 
the GSS envelope is more metal-poor than the GSS core \citep{ibata2007}.
A two-sided Kolmogorov-Smirnov (KS) test returns a probability of 
0.001\% that the MDF of the kinematically cold component in field a13 and 
the MDF of the GSS core were drawn from the same distribution.   
If we assume that the GSS envelope follows the same trend of 
increasing line-of-sight distance with \rproj\ as the GSS core 
and correct for this effect, the
difference in the metallicity between the GSS core in fields f207, H13s, and a3 and the GSS envelope in field a13 becomes slightly greater ($\sim 0.8$~dex). 
The largest systematic effect in this field may come from the 
identification of individual stars as part of the kinematically cold 
component.  The double-Gaussian fit to the stars in this field 
(Table~\ref{table:gss_mlfits}) implies that 13.5\% of the stars 
($\sim 5$ stars) within $\pm 2\sigma_v$ of
the mean velocity of the kinematically cold component are associated with 
the spheroid.  If we assume that the 5 lowest metallicity stars within 
$\pm 2\sigma_v$ of the kinematically cold component are M31 halo stars, 
the estimated \fehp\ of the GSS envelope debris increases by $+0.25$~dex ($\langle$[Fe/H]$\rangle_{\rm med}=-0.92$, $\langle$[Fe/H]$\rangle_{\rm mean}=-1.00\pm0.10$).

\subsection{Minor Axis Arc at \rproj\,$\sim 57$~kpc}\label{sec:prop_maxis}
We identified two kinematically cold components in field m4: one 
at $v=-355$~\kms\ which is relatively metal-rich, and one at
$v=-255$~\kms\ which is more metal-poor 
(\S\,\ref{sec:m4}, Fig.~\ref{fig:m4}).  
This field is located squarely on the 60~kpc minor-axis arc ``stream C '' 
(Fig.~\ref{fig:gss_roadmap}), identified as a 
photometric overdensity in starcount maps.  The M31 starcount map 
does show evidence of two overdensities in the 
region of stream C: a more prominent overdensity that is relatively 
metal-rich, and a less prominent overdensity that 
is relatively metal-poor \citep{chapman2008}.  
Both the relative contributions of the kinematially-cold components 
in field m4 (Table~\ref{table:gss_mlfits}) and 
their metallicities (Fig.~\ref{fig:m4}, discussed in detail below) 
imply that the 
$v\sim-355$~\kms\ component corresponds to the former, and 
the $v\sim-255$~\kms\ component corresponds to the latter.
 
Our field is 1.1\degree\ 
northeast of the stream C spectroscopic fields F25 and F26 presented 
by \citet{chapman2008}.  They also find two kinematically 
cold components in this region: a relatively metal-rich stream `Cr' 
[$\langle v \rangle=-349.5$~\kms\ and $\sigma_v^{\rm intr}=5.1\pm2.5$~\kms\ 
based on 9 stars (quoted errors are $1\sigma$)], 
and a relatively metal-poor stream `Cp' ($\langle v \rangle=-285.6$~\kms\ and 
$\sigma_v^{\rm intr}=4.3^{+1.7}_{-1.4}$~\kms\ based on 6 stars).  
Arrows in Figure~\ref{fig:m4}({\it a}) mark the mean line-of-sight 
velocities and intrinsic velocity dispersions of streams Cr and Cp in the 
F25/F26 field.   Due to the spatial proximity of our m4 spectroscopic masks 
and the \citet{chapman2008} F25/F26 masks as well as the proximity of the 
kinematical features in velocity space, it is reasonable to identify 
our $v=-355$~\kms\ component with stream `Cr' and our $v=-255$~\kms\ 
component with stream `Cp.'  A comparison of the mean velocities of 
the kinematically cold components 
in the m4 and F25/F26 fields suggests that their velocities diverge as 
one moves from the southwest to the northeast along stream C.  

The \ivio\ CMD distributions of stars within $\pm2\sigma_v$ of the mean 
velocities of the $v=-355$~\kms\ and $v=-255$~\kms\ kinematically cold 
components in field
m4 are shown in the bottom panel of Figure~\ref{fig:cmd_3pnl}.  It is
clear that on average the stars in the $v=-355$~\kms\ primary component
 are redder than the stars in the $v=-255$~\kms\ secondary component. 
The [Fe/H] distributions of these stars (assuming a coeval population
 of 12~Gyr) is displayed in Figure~\ref{fig:mdf_gss}({\it c}).  The     
differences seen in the \ivi\ CMD translate to clear differences in the
metallicity distribution functions of the stars.  While the [Fe/H] 
distribution of the $v=-355$~\kms\ primary cold component in
field m4 is relatively metal-rich and similar to the [Fe/H] distribution
 of the GSS, the MDF of the $v=-255$~\kms\ secondary cold component is 
significantly more metal-poor.

The median metallicity of the $v\sim-355$~\kms\ kinematically cold component 
is $\langle$[Fe/H]$\rangle_{\rm med}=-0.56$ 
($\langle$[Fe/H]$\rangle_{\rm mean}=-0.79\pm 0.12$), 
which is similar to the metallicity measured for 
stream C of [Fe/H]$=-0.6$ \citep[][measured from the region shown in 
Fig.~\ref{fig:gss_roadmap}]{ibata2007}.  Using 10~Gyr old Padova 
isochrones \citep{girardi2002}, \citet{chapman2008} 
find an average metallicity of $\langle$[Fe/H]$\rangle_{\rm avg}=-0.74\pm0.19$ 
for stream Cr.  

Based on the double-Gaussian fit to the velocities in this field 
(Table~\ref{table:gss_mlfits}), 14\% of the stars within $\pm 2\sigma_v$ of the $v\sim-355$~\kms\ kinematically cold component in field m4 are 
likely associated with M31's spheroid ($\sim 3$ stars).  M31's spheroid 
has been found to be metal-poor at large radii \citep{kalirai2006halo,koch2008}.  
If we assume that the lowest metallicity stars are those associated with 
M31's spheroid, the estimated median \fehp\ of the kinematically cold component
increases by only $+0.02$~dex. 

The $v\sim-255$~\kms\ kinematically cold component is on average 
more metal-poor than the $v\sim-355$~\kms\ component 
[Fig.~\ref{fig:mdf_gss}({\it c})].  The median 
metallicity of stars within $\pm 2\sigma_v$ of the mean velocity is 
$\langle$[Fe/H]$\rangle_{\rm med}=-1.33$ 
 ($\langle$[Fe/H]$\rangle_{\rm mean}=-1.31\pm 0.18$).  
Using 10~Gyr old Padova isochrones, \citet{chapman2008} find an 
average metallicity of 
$\langle$[Fe/H]$\rangle_{\rm mean}=-1.26\pm0.16$ for stream Cp.  
 
\paragraph{A physical association with the GSS?} 
Based on an analysis of all stars within the stream C and GSS core 
regions marked in Figure~\ref{fig:gss_roadmap}, \citet{ibata2007} 
claimed an association between the GSS and the 60~kpc minor-axis 
arc was unlikely due to significant differences in the properties 
of their stellar populations.  However, if the GSS's progenitor was 
a disk galaxy with a reasonable metallicity gradient, simulations show 
that arc-like features along M31's southeast minor axis with different 
stellar populations from the GSS can be produced during the progenitor's 
disruption \citep{fardal2008}, although they are farther from M31's 
center than the observed minor-axis arcs.  The simulated arc features 
have mean line-of-sight velocities that are close to the systemic 
velocity of M31 ($v_{\rm sys}=-300$~\kms).  There is also observational 
precedence for variations in metallicity along a tidal feature: a strong 
metallicity gradient has been observed in the Milky Way's Sagittarius 
stream \citep{chou2007}.  Therefore, it remains possible that the GSS and 
some of the substructure seen in this region share a common progenitor.  
Regardless, the relatively high metallicity of the $v=-355$~\kms, 'Cr'
stream indicates its progenitor was likely fairly massive 
\citep[e.g.,][]{larson1974,mateo1998,dekel2003}.

\paragraph{Connection with the extended cluster EC4} 
\citet{collins2009} find that stream Cp from the \citet{chapman2008} study 
has a line of sight velocity and metallicity consistent with
the extended cluster EC4, which was discovered as part of the CFHT/Megacam
survey of M31 \citep{ibata2007} and had RGB members which were targeted on the 
F25/F26 masks.  EC4 has a half-light radius of 33.7~pc and 
$M_V=6.6$ \citep{mackey2006}, placing
it solidly in the gulf between globular clusters and dwarf spheroidals in 
size-luminosity space.  One proposed explanation for such objects is that they
are the stripped cores of dSphs.  It is thus tempting to consider the 
possibility that stream Cp is the tidal stream of a stripped dSph, whose core
remains intact as the cluster EC4.  However, the measured M/L ratio for EC4 of 
$6.7^{+15}_{-6.7}$\Msun/\Lsun\ \citep{collins2009} makes this unlikely.  
\citet{collins2009} argue that it is 
also unlikely the stream Cp 
stars are tidally disrupted members of EC4 itself, since they are between
15\,--\,74 core radii distant from EC4 and HST photometry of EC4 shows no 
evidence of distorted isophotes or significant tidal debris 
\citep{mackey2006}.  Our discovery of 
the continuation of stream Cp in a field 1.1\degree\ (15~kpc) to the 
NE confirms that stream Cp is not formed from tidally stripped cluster stars.
It is most likely that stream Cp and the $v=-255$~\kms\ stream 
in field m4 represents the tidal stream of an unknown progenitor, 
from which the extended cluster EC4 was also tidally stripped. 

In conclusion, we find that the metallicity distribution 
of stars kinematically associated with the $v\sim-355$~\kms\ 
cold component and the GSS core are similar (Fig.~\ref{fig:mdf_gss}), 
while the metallicity distribution of stars associated with the 
$v\sim-255$~\kms\ is significantly more metal-poor than the GSS core.  
The two kinematically cold components in field m4 are consistent with being
detections further along stream C of the kinematical substructures 
`Cr' and `Cp' \citep{chapman2008}.  The progenitors of these streams 
remain unknown.  
   
\subsection{Second Kinematically Cold Component in the GSS Core Fields}\label{sec:prop_sec}

\subsubsection{Radial Trends}\label{sec:prop_sec_data}

\begin{figure}[tb]
\epsscale{1.0}
\plotone{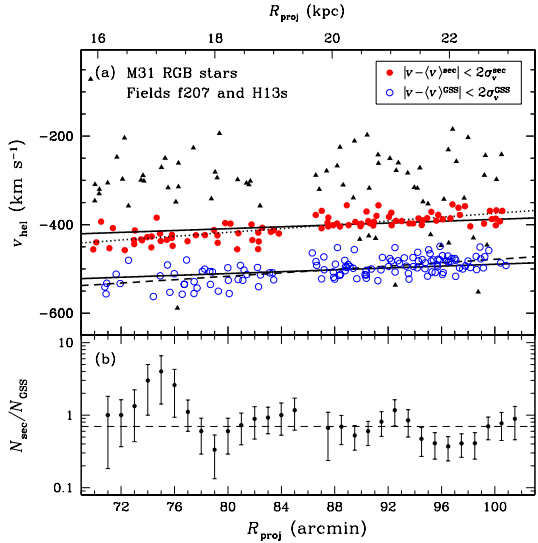}
\caption{Comparison of stars associated with the GSS core and the second kinematically cold component in fields f207 and H13s, at  \rproj\,$\sim 17$ and 21~kpc, respectively (\S\,\ref{sec:prop_sec_data}).  ({\it a}) The line-of-sight velocity distribution of M31 RGB stars in fields f207 and H13s as a function of projected radial distance from M31's center.  Open circles denote stars within $\pm 2\sigma_v$ of the mean velocity of the GSS core in each field, while solid circles denote stars within $\pm 2\sigma_v$ of the mean velocity of the second kinematically cold component in each field.  Solid triangles denote the remainder of the M31 RGB stars in these fields.  The solid line is the linear fit to the mean velocity of the GSS core as a function of radius (Fig.~\ref{fig:veltrends}, \S\,\ref{sec:prop_gss_kin}), offset by $+100$~\kms\ for the second kinematically cold component.  The dashed and dotted lines are the linear least-squares fits to the GSS core and second kinematically cold component data, respectively.  
({\it b}) The ratio of the number of stars in the second kinematically cold component and the GSS core ($N_{\rm sec}/N_{\rm GSS}$) as a function of radius.  The ratio was calculated using running bins of width 3\arcmin.  The error bars are based on Poisson statistics.  There is no apparent trend with radius in $N_{\rm sec}/N_{\rm GSS}$. The dashed line represents a constant ratio of $N_{\rm sec}/N_{\rm GSS}=0.70$, which is the ratio of stars in the second kinematically cold component to stars in the GSS based on the fits in Table~\ref{table:gss_mlfits}.
}
\label{fig:sec_comps_vel}
\end{figure}

Each of the two innermost GSS core fields (adjacent to each other at 
\rproj\,$\sim 17$ and 21 kpc) has a second kinematically cold component in 
addition to the primary kinematical component identified as the 
GSS core (Fig.~\ref{fig:gss_velhist}, 
\S\S\,\ref{sec:ongss_f207}\,--\,\ref{sec:ongss_H13s_a3}).
The top panel of Figure~\ref{fig:sec_comps_vel} displays line-of-sight 
velocity vs.\ \rproj\ for the M31 RGB stars in fields f207 and H13s.  
Stars within $\pm 2\sigma_v$ of the mean velocity of the GSS core and 
second kinematically cold component in each field 
(Table~\ref{table:gss_mlfits}) are denoted by blue and red points, 
respectively.  Stars comprising the main GSS peak and the second cold 
component follow nearly parallel linear trends in the  
line-of-sight velocity vs.\ \rproj\ plane.  These trends are continuous 
across the boundary between the two fields.

The second cold component, like the GSS, is evenly distributed across the 
two fields.  The bottom panel of Figure~\ref{fig:sec_comps_vel} shows the 
ratio of the number of stars in the second kinematically cold component to 
the number in the GSS ($N_{\rm sec}/N_{\rm GSS}$) as a function of \rproj.  
There is no apparent trend in the ratio $N_{\rm sec}/N_{\rm GSS}$ with \rproj.   

A more accurate measure of the velocity dispersions of the two kinematically cold
components can be obtained by accounting for the linear trend of line-of-sight 
velocity with \rproj\ (Fig~\ref{fig:sec_comps_vel}).   The line-of-sight velocities of 
individual stars are shifted to a reference frame defined by the line-of-sight velocity 
of the GSS as a function of \rproj.  A maximum-likelihood, triple-Gaussian fit to the data 
(\S\,\ref{sec:cks}) is performed.  The best-fit values for the observed velocity
dispersions of the GSS and second kinematically cold component are 
$20.1^{+3.8}_{-3.1}$~\kms\ 
and $14.1^{+7.1}_{-6.0}$~\kms, respectively.  These measurements correspond to intrinsic 
velocity dispersions of 19.4 and 13.0~\kms, respectively, when the mean radial velocity 
measurement error in each field is taken into account (\S\S\,\ref{sec:ongss_f207}\,--\,\ref{sec:ongss_H13s_a3}).

There is no evidence in the other GSS core fields of a continuation of the 
second kinematically cold component seen in fields f207 and H13s 
(Fig.~\ref{fig:gss_velhist}).  If the linear trend seen in 
Figure~\ref{fig:sec_comps_vel} is extrapolated, the mean velocities expected 
for the feature in fields IF8 (\rproj\,$\sim 12$~kpc), IF6 
(\rproj\,$\sim 26$~kpc), and a3 (\rproj\,$\sim 33$~kpc) are $-476$, $-344$ 
and $-278$~\kms, respectively.  The field d1 does have a second kinematically 
cold component, which is also offset by $\sim100$~\kms\ from the primary (GSS core) 
kinematically cold component.  It is possible that the second kinematically
cold component in fields f207 and H13s is a tidal stream that follows a radial line
from fields f207 and H13s to field d1 (and therefore is not present in fields IF6 and a3).
However,  this interpretation is unlikely based on the metallicities of the stars.
Stars in the secondary kinematically cold component
in field d1 are uniformly metal-rich (Fig.~\ref{fig:vf_and1}), and are significantly more metal-rich than the second kinematically cold component
in fields f207 and H13s (\fehp$\sim -0.2$~dex, compared to \fehp$\sim -0.5$~dex).

\subsubsection{Implications for the Origin of the Second Kinematically Cold Component}\label{sec:prop_sec_interp}
The discovery of the second kinematically cold component ($\langle v \rangle = -388$~\kms) in the GSS core field at \rproj\,$\sim 21$~kpc, field H13s, was originally reported by \citet{kalirai2006gss}.  These authors considered three possible origins for the secondary kinematically cold component: (i) a feature related to M31's disk, (ii) a separate halo tidal stream superimposed on the GSS, and (iii) a wrapped around component of the GSS.  In this section, we re-evaluate the merits of each of these as the possible physical origin of this feature in light of the newly obtained data in the GSS core field f207 (\rproj\,$\sim 17$~kpc), the improved statistics in field H13s (\S\,\ref{sec:prev_pub}), and recent theoretical and observational developments.  We also consider a fourth possible origin which was not considered by \citet{kalirai2006gss}: a bifurcation in velocity-space of the GSS.  We will argue that the data in fields f207 and H13s favor 
a physical connection with the GSS rather than a separate physical origin for the 
second kinematically cold component in fields f207 and H13s. 

\paragraph{M31's stellar disk} The disk of M31 has been observed to 
large radial distances from the center of M31.  An extended smooth 
stellar distribution has been observed out to $\sim 40$~kpc in the 
disk plane, and substructures with disk-like velocities have been observed
continuing out in radius to 70~kpc \citep[e.g.,][]{ibata2005}.  
Furthermore, the mean line-of-sight velocity of the second
kinematically cold component is within $\sim 50$~\kms\ of the expected
line-of-sight velocity for M31 disk stars in circular orbits at the location
of our masks.  We 
therefore consider the possibility that the secondary kinematically 
cold component in fields f207 and H13s represents a population of smooth 
disk stars.  \citet{kalirai2006gss} show that the surface brightness of 
the secondary kinematically cold component in field H13s 
[$R_{\rm disk}\sim 75$~kpc assuming an unwarped disk of inclination 
77\degr\ \citep{walterbos1988}] is more than 25 times brighter than would 
be expected based on an extrapolation of M31's smooth exponential disk, 
effectively ruling out the smooth disk of M31 as the origin of the 
substructure \citep[Fig.~11 of][]{kalirai2006gss}.  The surface brightness 
of the secondary peak in field f207 ($R_{\rm disk}\sim 59$~kpc) is 
equivalent to that of the secondary kinematically cold component in field 
H13s, and is approximately 10 times brighter than the extrapolation of 
M31's smooth disk.  The surface brightness of M31's disk 
decreases exponentially with increasing projected radius while the 
surface brightness of the GSS is relatively constant over the radial 
range covered by the f207 and H13s spectroscopic masks.  The
f207 and H13s masks combined cover a distance of $\sim 30$~kpc in the plane 
of the disk, which is more than 5 scale lengths \citep{walterbos1988}. 
Thus, the ratio of the number of stars in the second kinematically 
cold component 
to the number of stars in the GSS would be expected to decrease 
substantially with increasing projected radius if the second 
kinematically cold component was part of M31's disk.  This is contrary to 
what is observed [Fig.~\ref{fig:sec_comps_vel}({\it b})].  
These arguments rule out the possibility that the origin of the second
kinematically cold component in these fields is M31's extended stellar disk.

\paragraph{Unrelated substructure} Cosmological simulations of halo 
formation \citep[e.g.,][]{bullock2001,bullock2005} as well as observational 
studies of the Milky Way 
\citep[e.g.,][]{majewski1996,newberg2002,belokurov2006} suggest that 
multiple tidal streams may exist along a single line of sight.  The 
secondary kinematical peaks in fields f207 and H13s could represent a tidal stream 
in M31's halo unrelated to the progenitor of the GSS.  Also, as mentioned 
above, substructures rotating with disk-like velocities have been 
observed as far as 
$\sim 70$~kpc from M31's center \citep{ibata2005}.  The arguments 
against a smooth disk origin for the second kinematically cold component 
in fields f207 and H13s do not preclude the possibility that it
is substructure related to the disk.  However, 
for either halo or disk-related substructures we would not expect a 
direct correlation between the line-of-sight velocity of the GSS and 
the second stream. 
Since the two kinematical substructures are superimposed on each other
in projected location, only a limited set of orbits would
produce the observed correspondence in mean line-of-sight velocity 
with \rproj\ (Fig.~\ref{fig:sec_comps_vel}). 

\paragraph{Wrapped around GSS component} \citet{kalirai2006gss} discussed the possibility 
that the second kinematically cold component in field H13s represented part of a later orbital phase of the GSS's progenitor that has wrapped around M31 and back to the location of this field.  A few of the many dynamical models of the merging event 
\citep{ibata2004,font2006gss,fardal2006} yield a progenitor orbit that passes through the region of fields H13s and f207 twice \citep[e.g., Fig.~8 of][]{fardal2006}.  This origin was disfavored by \citet{kalirai2006gss} because the stars are predicted to be moving in very different phases during the two passages.  This results in a predicted line-of-sight velocity offset between the stellar streams produced by each phase that is inconsistent with the observed velocity offset between the GSS core and the second kinematically cold component in fields H13s and f207 (Table~\ref{table:gss_mlfits}).  
The most recent dynamical models of the collision suggest that the Northeast and Western shelves observed in M31 are the forward continuation of the GSS \citep{fardal2007}.  This model is strongly supported by recent spectroscopic and HST/ACS observations of the shelves in M31 \citep{brown2006apjl,gilbert2007,richardson2008}.  Although in this model the progenitor's orbit does not pass through the region of fields H13s and f207 a second time, the shelves themselves could be responsible for the second kinematically cold component.  The extent of the W shelf is obscured by the southwest region of M31's disk.  It could produce a velocity peak with a $\sim 100$~\kms\ separation from the GSS velocity in fields f207 and H13s, although the second velocity peak characteristic of shelf features (one peak corresponds to outflowing and one to infalling debris) would have to be depopulated to match the data  (M. Fardal, private communicaiton).  The W shelf would have to span almost 180\degree\ in position angle to be the source of the kinematically cold component in fields f207 and H13s.    Future spectroscopic observations will test whether the velocity of W shelf stars are consistent with the second kinematically cold component in the GSS core fields f207 and H13s.

\paragraph{Bifurcation of the GSS} The fact that the velocities of stars 
in the GSS and second kinematically cold component follow the same linear 
trend with projected distance from M31's center, and that the ratio of 
stars in the two kinematical features is approximately constant with \rproj, 
suggest the possibility of a direct physical association between the two 
features.  The formation of velocity caustics during the tidal disruption 
of an in-falling satellite can lead to bifurcations in the observed 
line-of-sight velocities of a tidal feature 
\citep[e.g.,][]{merrifield1998,fardal2007}.  Thus, the second kinematically 
cold component in fields f207 and H13s could be part of the GSS itself 
(i.e., stars, like those in the GSS, that are approaching their first 
pericentric passage since the progenitor's initial encounter with M31).  
However, it may be difficult to produce a bifurcation with a separation of 
line-of-sight velocities as large as $\sim 100$~\kms, as seen in fields f207 and
H13s.  Further modeling of the collision 
will be needed to determine whether velocity caustics of this magnitude 
can be produced in this region of the GSS. 

\begin{figure}[tb]
\plotone{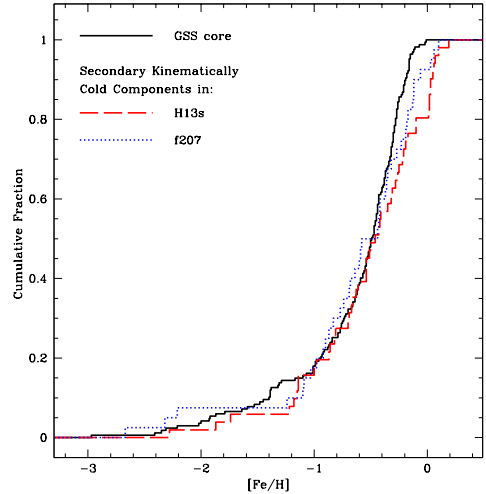}
\caption{Cumulative [Fe/H] distribution of stars within $\pm2\sigma_v$ of the mean velocity of the second kinematically cold components in fields H13s (dashed curve) and f207 (dotted curve).  The cumulative [Fe/H] distribution of the GSS core (based on the primary kinematically cold component in fields f207 and H13s and the kinematically cold component in the a3 field) is shown for comparison (solid curve; Fig.~\ref{fig:mdf_gss}).  The second kinematically cold component in fields f207 and H13s is as metal-rich as the GSS core. 
}
\label{fig:sec_comps_met}
\end{figure}

If the second kinematically cold component is related to the GSS, either as
a bifurcation of the GSS itself or the forward continuation of the GSS, 
the metallicity distributions of stars in the GSS and the second 
kinematically cold components should be similar.  
Figure~\ref{fig:sec_comps_met} displays the cumulative [Fe/H] distributions 
of stars in the GSS core (solid curve) and the second kinematically cold 
component in field f207 (dotted curve) and H13s (dashed curve).  Although 
the second kinematically cold components have a small number of stars more 
metal-rich than any associated with the GSS core, the [Fe/H] distributions 
are not inconsistent.  A two-sided KS test returns probabilities of 15\% 
and 22\% that the [Fe/H] distributions of the second kinematically cold 
components in f207 and H13s, respectively, were drawn from the same parent 
distribution the GSS core was drawn from.  

In conclusion, the data rule out M31's extended, smooth disk as the 
origin of the second kinematically cold components in fields f207 and H13s.  
While substructure in M31 unrelated to the GSS cannot be ruled out by
the data, the possibility of a connection with the GSS, either via a bifurcation in 
velocity space of the GSS itself or a wrapped around component of the GSS, 
is an intriguing possibility.  The common trend of 
mean line-of-sight velocity with projected radius of the GSS and the 
second kinematically cold component combined with the similar metallicity 
distributions of their stars would be a remarkable coincidence if the 
substructures are unrelated, but are naturally explained if the second 
component is also part of the GSS.       

\subsection{Summary of Debris Properties}
We have detected the metal-rich GSS core
in fields f207, H13s, a3, and d1.  Although field d1 is located on
the dSph And~I, we are still able to isolate M31 field stars from dSph members,
and to identify kinematically cold features in the M31 field star population.
The kinematical and chemical properties of the $-380$~\kms\ kinematically cold
component in field d1 are entirely consistent with its identification as 
GSS core debris.  The kinematical and chemical properties of the 
relatively metal-poor $-300$~\kms\ cold component in field a13 
are consistent with what is expected for the GSS envelope 
based on empirical evidence from photometric studies and theoretical
modeling of the collision that produced the GSS.   
The second kinematically cold component 
in fields f207 and H13s has a similar \fehp\ distribution to the GSS core, 
follows the same parallel trend in the line-of-sight velocity vs.\ \rproj\ 
plane, and may also be associated with the GSS.

The kinematically cold features observed in
field m4 confirm the identification by \citet{chapman2008} of two 
kinematically cold components (one metal-rich and one metal-poor) in
the photometric stream C minor-axis feature. The origin of the more metal-rich
kinematical feature is not known.  The more metal-poor feature may be 
physically associated with the extended stellar cluster EC4, although it is 
unlikely to be a tidal stream originating from EC4 itself.  

\section{Discussion}\label{sec:discussion}

Mixed stellar populations appear to be an inevitable feature of the M31 halo.  
Many fields in the SPLASH spectroscopic survey show conclusive (Table~\ref{table:gss_mlfits}) or 
suggestive evidence of multiple components within the 
M31 field halo population itself.  In addition, fields that target M31's dSph galaxies 
contain dSph member RGB stars as well as M31 field halo RGB stars (\S\,\ref{sec:and_fields}).  
All fields contain a non-negligible fraction of foreground Milky Way dwarf star 
and background galaxy contaminants.   It is useful to keep the notion of mixed populations in
mind while interpreting star-count maps.
  In this section we discuss several implications of the mixed stellar populations discovered in M31 for our understanding of the structure and formation of M31's stellar halo.  
 
Although many of the spectroscopic fields presented here specifically targeted tidal debris (f207,  H13s, IF8, IF6, IF2, and IF1), others were chosen for DEIMOS spectroscopy without prior knowledge of the existence of tidal debris at their locations (fields a3, m4, a13, d1, and d3).  Nevertheless, the majority of stars are found to be located in distinct clumps in $v_{\rm los}$\,--\,[Fe/H] space.  Moreover, many fields contain multiple distinct groupings of stars.  If the outer regions of M31's stellar halo were well-mixed,  stars from numerous progenitors would be present along any given line of sight, leading to stars that are broadly distributed in $v_{\rm los}$\,--\,[Fe/H] space.  \citet{tremaine1993} calculates that the filling factor of tidal streams, $f_{3}$, in a Milky-Way type halo built from the disruption of accreted satellites should be small ($f_{3}< 1$) at radii of $\sim50$~kpc (a given volume of the halo is dominated by stars from one to a few progenitors), while well-mixed systems have $f_{3}>\!\!>1$ (stars from numerous progenitors are present in any given volume).  The fields presented above indicate that a large fraction of the stars in M31's halo at these radii belong to groups of stars that can be identified as individual tidal streams.  As discussed by \citet{majewski1996}, this may prove problematic for drawing conclusions about the global properties of M31's stellar halo based on limited lines-of-sight.    
Further studies in other quadrants of M31's stellar halo will be able to determine whether the dominance of distinct groups of stars in these fields is a ubiquitous property of M31's stellar halo, or is mainly a result of the large amounts of debris deposited by the progenitor of the GSS.    

The number of distinct kinematical components present in many of the M31 stellar halo fields necessitates   
careful interpretation of the properties of specific regions and tidal features measured from photometric studies of M31's halo population.  
Even if the percentage of the population associated with a tidal feature is known via separate kinematical studies, photometric studies 
cannot identify which stars are more likely associated with a specific feature rather than the underlying halo of M31.  Metallicity distributions of features derived purely from photometric observations will therefore be a sum of the metallicity distribution of the feature itself, M31's halo, and any other structural (sub)component or feature present in the analyzed region.  The minor-axis arc stream C provides a prime example of the complexity that can be present in a small region of M31's halo (\S\,\ref{sec:m4} \& \S\,\ref{sec:prop_maxis}).  Furthermore, the extensive presence of mixed stellar populations in M31's stellar halo, seen in both kinematical \citep[this work; also see][]{ibata2004,ibata2005,guhathakurta2006,kalirai2006gss,gilbert2007} and photometric surveys \citep[e.g.,][]{ibata2001nature,ferguson2002,ibata2007}, make identifying control M31 halo fields for photometric studies problematic.  Kinematical data provide the ability to identify which stars are likely to be associated with an individual (sub)component or tidal feature; this provides a distinct advantage over photometric studies in the analysis of the properties of individual stellar (sub)components in M31.
  
A variety of spatial structures are expected in stellar halos that form via the merging of smaller stellar systems, including still-bound satellites, spatially coherent tidal streams, larger, more diffuse clouds of debris, and a spatially smooth, well-mixed stellar population. The data presented here display evidence of a correlation between kinematical and spatial structure.  The M31 dSphs And~I and And~III are very kinematically cold, with low velocity dispersions ($\sigma_v < 10$~\kms; Kalirai et al.\ in preparation).  As surviving dSphs, these objects are also spatially compact.  Many of the kinematically cold components observed in our spectroscopic fields (i.e., m4 and the GSS core in f207, H13s, and a3) can be identified with spatially sharp debris streams.  The kinematically cold component in field a13 is warmer (has a broader velocity distribution) than the kinematically cold components in the GSS core fields.  It is identified with the GSS envelope debris, which appears to be a broad cloud of debris rather than a sharp tidal stream.  Our fields also contain a kinematically hot population with a very broad spread in velocities; this population represents M31's underlying, well-mixed stellar halo.  

Metal-rich tidal streams are expected to dominate over metal-poor streams in surface brightness limited stellar halo surveys, due primarily to the fact that more massive dwarf galaxies are both more metal-rich \citep[e.g.,][]{larson1974,mateo1998,dekel2003} and tend to produce higher surface brightness stellar streams \citep{johnston2008,gilbert2009metrich}.  
We currently have $\lesssim 50$~M31 RGB stars per spectroscopic field in M31's outer halo (\rproj\,$\gtrsim 35$~kpc).  A high surface brightness tidal stream is easily identifiable in our survey in the form of a kinematically cold component seen against the dynamically hot underlying stellar halo.
Tidal streams with a low level of contrast against M31's spheroid are difficult to detect due to our limited sample size.  We therefore expect the kinematically cold components identified in our M31 fields to be 
more prominent among the metal-rich 
than metal-poor 
stars.  In general this is indeed the case, as shown in panels ({\it b--d}) of Figs.~\ref{fig:f207}\,--\,\ref{fig:m4}.

\section{Conclusions}\label{sec:gss_concl}
In this study, we presented data for five 
new spectroscopic fields observed
during the course of our SPLASH Keck/DEIMOS survey.  We also presented 
improved reductions and sample selection for two of our previously published fields 
\citep[a3 and H13s;][]{guhathakurta2006,kalirai2006gss} and compiled all previously
published line-of-sight velocity data in fields on the GSS.   This combined spectroscopic 
dataset was used to analyze the kinematical and chemical properties of the GSS.

The main findings of this paper are:
\begin{itemize}
\item{We have detected and measured the mean velocity and velocity dispersion of the GSS in two new GSS core fields:  (i) field f207, at \rproj$\sim17$~kpc (the innermost kinematic measurement of the GSS to date), and (ii) field d1 at \rproj$\sim45$~kpc (the first detection of GSS core debris in a field well removed from the sharp eastern edge of the GSS).}
\item{We have made the first kinematical detection and measurement of the mean velocity and velocity dispersion of the GSS envelope in field a13, at \rproj$\sim 58$~kpc.}
\item{We have presented updated measurements of the mean velocity and velocity dispersion of two previously published GSS core fields at \rproj$\sim 21$ and 33~kpc using improved reductions of the spectroscopic data.}
\item{The full dataset of new and previously published line-of-sight velocities in GSS fields was used to measure the dependence of the mean line-of-sight velocity and velocity dispersion of the GSS as a function of projected distance from M31's center.} 
\item{We utilized the line-of-sight velocities in the SPLASH survey fields to identify stars belonging to the GSS.  The metallicity distribution of stars in the GSS is found to be similar in all GSS core fields, while the mean metallicity of stars in the GSS envelope is significantly more metal-poor ($\sim 0.7$~dex) than the GSS core.}
\item{We have confirmed the presence of two kinematically cold components, one relatively metal-rich and one relatively metal-poor, in the stream C region at \rproj$\sim 60$~kpc on the minor axis.  A comparison of the mean velocities of the kinematic features in field m4 with a previously published field 1\degree\ to the southwest indicates that the observed line-of-sight velocities of the two features diverge as one moves from the southwest to the northeast along stream C.}
\item{In our innermost GSS core field (\rproj$\sim 17$~kpc) we discovered the continuation of a second kinematically cold component previously seen in a GSS core field at \rproj$\sim 21$~kpc.  The kinematical and chemical properties and number of stars in this second kinematically cold component are tightly correlated with those of the GSS over $\Delta R_{\rm proj} =7$~kpc, indicating that a direct physical association between it and the GSS is likely.  The second kinematically cold component in these two fields may represent a bifurcation in the 
line-of-sight velocities of stars in the GSS.}
\end{itemize}

Previous spectroscopic data on the GSS has been presented piecemeal in 
several different papers \citep{ibata2004,guhathakurta2006,kalirai2006gss}.  
The compilation of previously published data and the new GSS fields presented here
provide a clear portrait of the kinematical and chemical properties of the GSS along
a significant portion of both its length and breadth.   This data provides 
ingredients for improved detailed modeling of the collision of the progenitor of the GSS 
with M31, and will provide 
additional observational constraints on M31's potential as well as the properties and orbit of the 
progenitor of the GSS.

\acknowledgments 
The authors thank Mark Fardal for many useful discussions.  This project 
was supported by NSF grants AST-0307966,
AST-0507483, and AST-0607852 (K.M.G., P.G.,
J.S.K., P.K., and E.N.K.), NSF Graduate Student Research Fellowships (K.M.G., P.K., and E.N.K.), 
and NSF grants AST-0307842,
AST-0307851 and AST-0607726, NASA/JPL contract 1228235, the David and Lucile Packard
Foundation, and The F.~H.~Levinson Fund of the Peninsula Community Foundation
(S.R.M., R.J.P., and R.L.B.).
J.S.K. was supported by NASA
through Hubble Fellowship grant HF-01185.01-A, awarded by the Space Telescope
Science Institute, which is operated by the Association of Universities for
Research in Astronomy, Incorporated, under NASA contract NAS5-26555.


\bibliography{m31}

\end{document}